\documentclass[aps,pre,floatfix,twocolumn,showpacs,superscriptaddress]{revtex4-1}
\usepackage{color,soul}
\usepackage{graphicx}
\usepackage{epstopdf}
\usepackage{amsmath, amsthm, amssymb}
\usepackage{amsthm}
\usepackage{multirow}
\usepackage[normalem]{ulem}
\usepackage{soul}

\newcommand{\be}{\begin{equation}}
\newcommand{\ee}{\end{equation}}
\newcommand{\bea}{\begin{eqnarray}}
\newcommand{\eea}{\end{eqnarray}}

\begin{document}
\title{On role of matrix behavior in compressive fracture of bovine cortical bone}
\author{Ashwij Mayya}
\email{ashwijmayya@gmail.com}
\author{Anuradha Banerjee}
\email{anuban@iitm.ac.in}
\affiliation{Department of Applied Mechanics, Indian Institute of Technology-Madras, Chennai 600036, India
}
\author{R. Rajesh}
\email{rrajesh@imsc.res.in}
\affiliation{The Institute of Mathematical Sciences, C.I.T. Campus, Taramani, Chennai 600113, India}
\affiliation{Homi Bhabha National Institute, Training School Complex, Anushakti Nagar, Mumbai 400094, India}
\date{\today}
\begin{abstract}
In compressive fracture of dry plexiform bone, we examine the individual roles of overall mean porosity, the connectivity of the porosity network, and the elastic as well as the failure properties of the non-porous matrix, using a random spring network model. Porosity network structure is shown to reduce the compressive strength by upto 30\%. However, the load bearing capacity increases with increase in
either of the matrix properties -- elastic modulus or failure strain threshold. To validate the porosity-based RSNM model with available experimental data, bone-specific failure strain thresholds for the ideal matrix of similar elastic properties were estimated to be within 60\% of each other. Further, we observe the avalanche size exponents to be independent of the bone-dependent parameters  as well as the structure of the porosity network.

\end{abstract}
\maketitle

\section{Introduction}
Cortical or compact bone, found in the mid-shaft of load bearing bones like femur and tibia, is a brittle, porous biomaterial. Being a living tissue, the local microstructure and porosity network of the bone evolves in response to the mechanical stresses that the bone is subjected to, and this in turn modifies the local mechanical properties. Understanding the relationship between microstructure and mechanical properties is crucial for applications such as extraction of bone grafts~\citep{pearce,conward2016}, in design of mechanically compatible implants~\citep{hing,bansiddhi,libonati2016} and porous scaffolds for bone tissue engineering~\cite{chen,liu}, to interpret loading history~\citep{keenan2017,zeddadifferences,mitchell2017}, to evaluate the effectiveness of chemical and physical therapeutical measures for bone healing~\citep{claes,gocha2015} etc. An important aspect of this understanding is the development and testing of models that incorporate microstructural features and predict material properties such as failure strength, elastic modulus, fracture paths, etc. Such models, if general enough, would also be  of use in understanding failure behavior of a wider class of brittle materials with a well defined porosity network, such as wood, rock ~\citep{meyers,shojaei} etc.

Compared to the detailed experimental characterization of the microstructure-property relationship of the different microstructures ~\citep{reilly,katz_1,katz_2,shahar,martin_1,martin_2,norman,currey_2,kim2005,niinomi,carter,li2013,scirep,msec}, the number of predictive microscopic models that are able to reproduce characteristic features of the complex fracture processes involved are few in number. This is primarily because the fracture process in complex heterogeneous quasi-brittle materials like bone involves multiple pores and micro-cracks that interact and evolve stochastically prior to final failure~\citep{bazant2004}, which is difficult to capture using deterministic models. Thus, modeling such materials using classical fracture mechanics theory or standard finite element method restricts the scope mostly to finding effective elastic behavior or to finding the resistance to growth of a single macroscopic crack~\cite{hoc,wahab,brown2014}. Statistical models, like the random spring network model (RSNM), that approximate the continuum by a network of springs with statistically distributed characteristics, are much better suited for studying fracture in such systems. RSNM has been successful in providing an insight into the role of disorder in the fracture behavior of heterogeneous material systems with no preexisting crack~\citep{alava}, reproducing features like transition from brittle to non-brittle macroscopic response~\citep{curtin}, avalanche size distributions~\citep{ray} and qualitative~\citep{moukarzel, urabe}  and quantitative~\citep{dhatreyi}  features of fracture of composite materials. In the context of bone, simple one-dimensional models of parallel springs with statistically distributed properties have been used to simulate the characteristic quasi-brittle softening seen in tension and bending~\citep{knovic} as well as in compression of cortical bone~\citep{schwiedrzik}. Including spatial effects, three-dimensional RSNM has been used to model cancellous bone, spongier bone found near joints, to account for percolation effects in the power-law variation of strength with porosity~\citep{gunaratne,rajapakse}. However none of these studies take into account the role of the structure of the porosity network. This is of particular importance since the structure of the porosity network, and not merely the mean porosity, is known to be important. Microstructures with larger mean porosity have sometimes higher compressive strengths than those with lower porosity~\citep{msec,li2013,trebacz}.

Cortical bone has predominantly two distinct microstructures -- plexiform bone that is brick shaped and has woven bone and vasculature sandwiched within regular lamellae, and Haversian bone that has cylindrical secondary osteons that run along the length of the bone~\citep{currey_book}. Under compression it has been found that mechanisms of failure are noticeably different between the microstructures. In plexiform bone damage is localized on weak radial planes resulting in prismatic fracture surfaces while in Haversian bone the presence of osteons deflects the crack paths and leads to meandering crack paths~\citep{msec}. 

To model the splitting fracture of plexiform bone under compression, recently we developed a 
 RSNM model that incorporates the details of the porosity network of the bone~\citep{jrsoc}. In our porosity-based RSNM (see Sec. II for more details), the local spring constants were determined based on the experimentally obtained porosity network of the plexiform bone. The effect of porosity network on the macroscopic compressive response of the bones was examined under the simplifying assumption that the material properties like Young's modulus, Poisson's ratio, and compressive strength are independent of the bone. While the model reproduced the overall force deflection curves, qualitative fracture paths, and avalanche exponents reasonably well, the correlation between the predicted and observed compressive strengths was not good. However, material properties are expected to be dependent on the bone, as well as the history of loading, and the discrepancy between model predictions and experimental data was attributed to ignoring bone dependent material properties. In this paper we introduce bone dependent material properties into the model by performing a detailed parametric study of the model by varying the model parameters systematically. From experimental energy dispersive spectroscopy, we find that elastic modulus is more or less independent of the bone. However, to obtain the experimental macroscopic response, we have to use sample-dependent strain thresholds. We conclude that samples with similar mean porosity and mineralization perhaps have different material organization, leading to variation in load bearing capacity. Further, the role of porosity network is examined by comparing the predictions with results from homogenized distribution of equivalent porosities. While homogenization leads to alteration of the failure paths and strengths, it leaves the avalanche exponents largely unchanged.

The remainder of the paper is organized as follows. In Sec. II, we describe in detail the model and how the porosity based RSNM is obtained from the experimental CT scan images. In Sec. III, results from parametric studies of the model parameters are discussed in terms of their effect on failure paths and load bearing capacity. Bone-specific properties are iteratively estimated to match with experiments and predictions are compared with simulations using bone-independent parameters. The exponents of avalanche size distributions are also presented for networked and homogenized porosity.

\section{Model}
\begin{figure}[h!]
\centering
\includegraphics[width = 7cm]{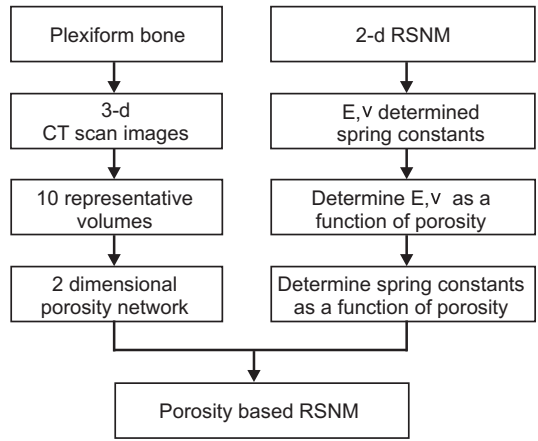}
\caption {Flow chart detailing the steps involved in developing porosity-based RSNM.}
\label{fig:flowchart}
\end{figure}

In this section, we describe  the formulation of the two dimensional spring network model for plexiform bone~\citep{jrsoc}, and the generalizations that will be studied in this paper. The incorporation of the experimentally obtained CT scan images into a porosity based two dimensional RSNM involves several steps that are summarized in the flowchart shown in Fig.~\ref{fig:flowchart}. We describe each of the steps below.

\begin{figure}[h!]
\centering
\includegraphics[width = \columnwidth]{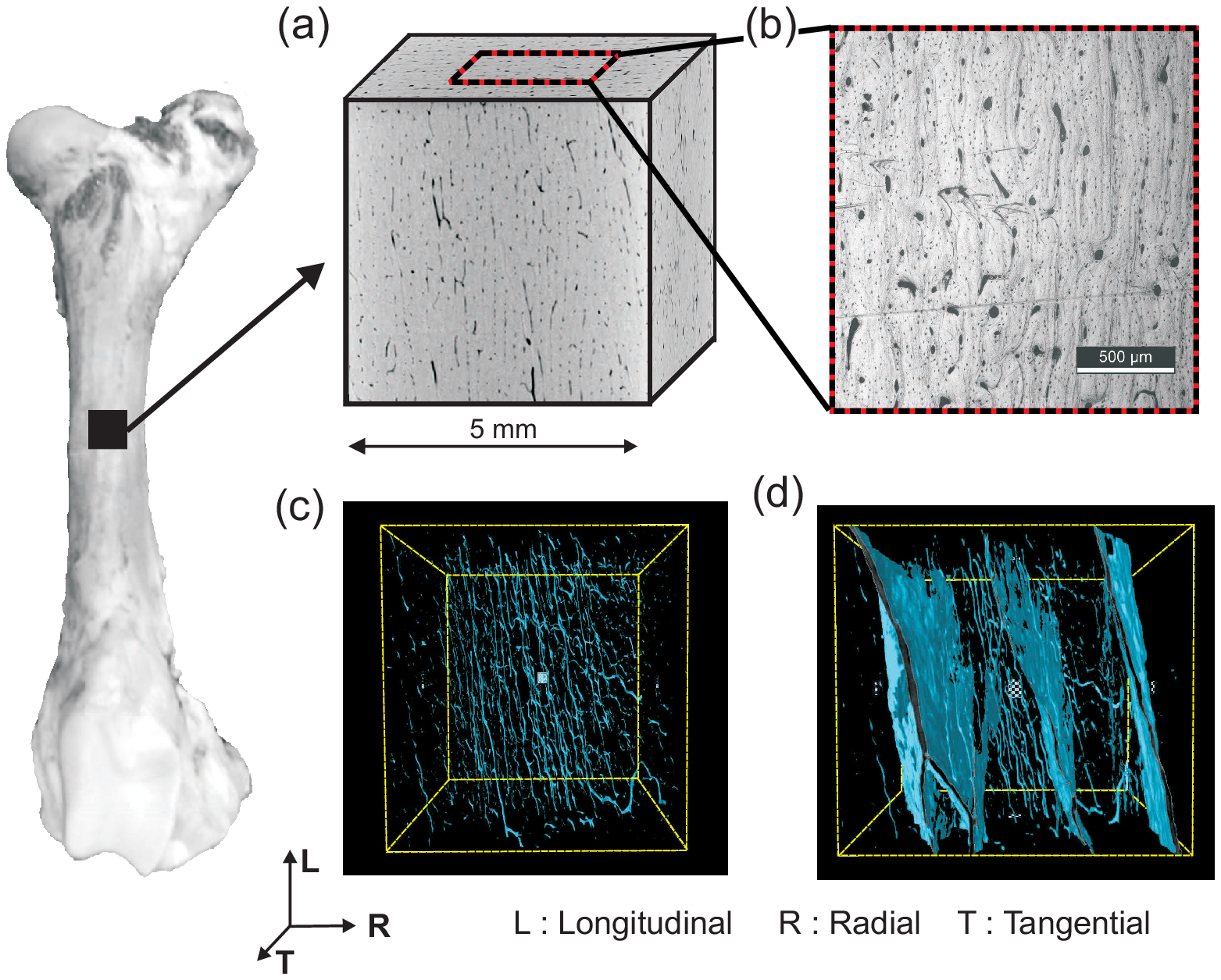}
\caption {\label{fig:setup} (a) Cubic sample of size $5 \mathrm{mm}\times 5 \mathrm{mm}\times 5 \mathrm{mm}$ with faces perpendicular to the longitudinal, radial and transverse directions. (b) Optical micrograph of the longitudinal face showing plexiform bone. CT scan images (tangential view) of the specimen (c) before fracture and (d) after fracture,  where blue points indicate higher porosity. Data are for sample III.}
\end{figure}
\begin{figure}[h!]
\centering
\includegraphics[width = \columnwidth]{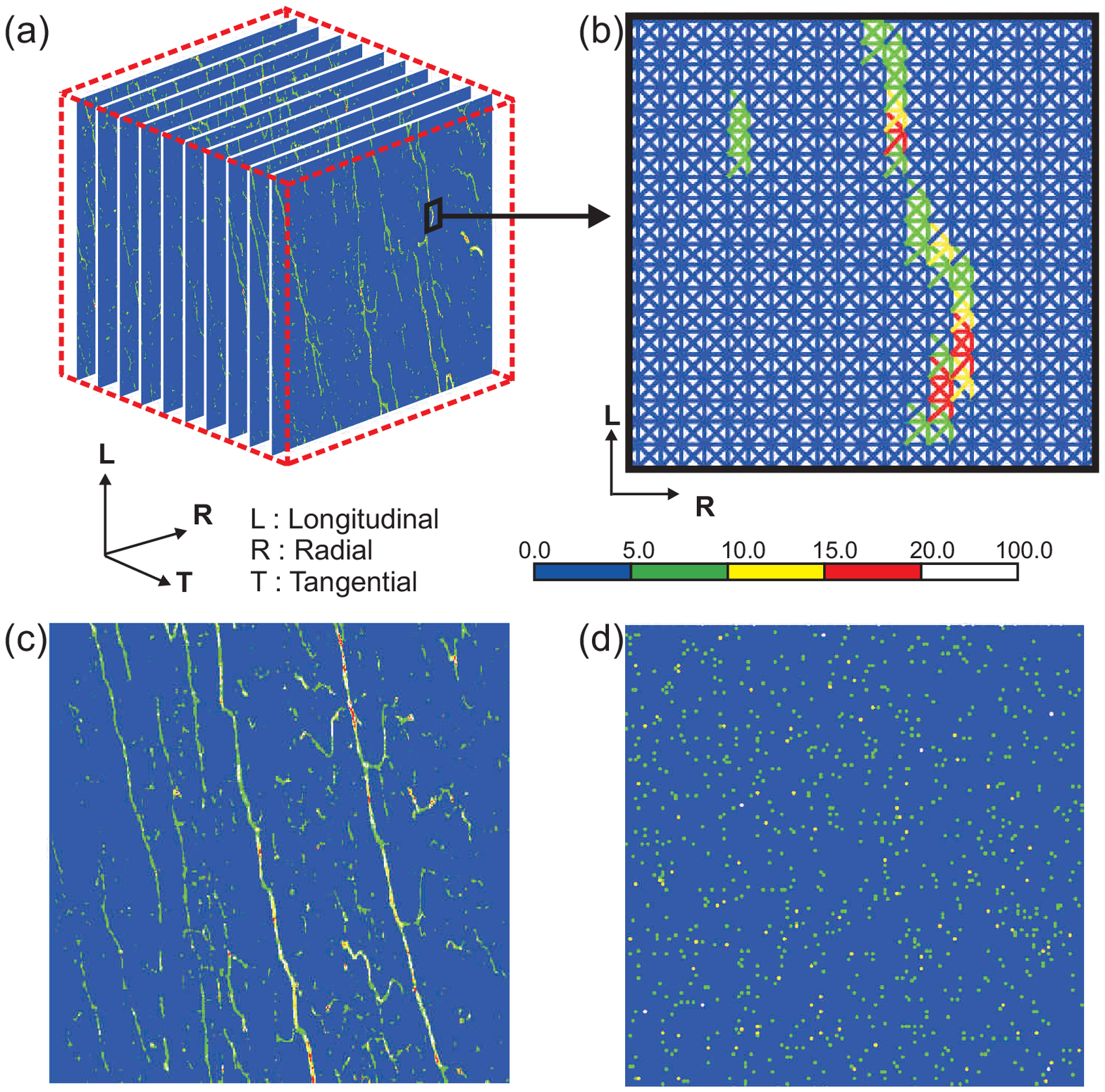}
\caption {\label{fig:representative domain} (a) Stack of representative domains, each spanning over $500~\mu m$ obtained from porosity analysis. (b) Spring stiffness are based on the \%porosity at the node (pixel). (c) A typical representative domain and (d) corresponding homogenized configuration for sample III.}
\end{figure}
We use the experimental data, reported earlier in \citet{jrsoc} in which cubical samples from the anterior section of the mid-diaphysis of bovine femur [see Fig.~\ref{fig:setup}(a)] were tested in compression along the length of the bone. CT scan images of the samples were obtained before and after compression failure~\citep{jrsoc}. Sample CT scan images of the pre- and post-compression samples are reproduced in Fig.~\ref{fig:setup}(c) and (d). The samples were chosen from regions that have plexiform microstructure, as evident from the brick-shaped layered structure seen in the optical micrograph shown in Fig.~\ref{fig:setup}(b).
The CT scan produces 500 slices, each slice corresponding to $10~\mu m$. The grayscale image  in each slice is converted to a binary image (0 or 1) by setting a threshold determined from the valleys of the probability distribution function of the grayscale. Here $1$ corresponds to a pore and $0$ to a non-pore. The 500 slices are now divided into 10 representative volumes, each consisting of 50 slices, corresponding to $500~\mu m$ thickness [see Fig.~\ref{fig:representative domain}(a)]. The thickness was chosen such that it is much larger than the mean pore size but less than the inter pore distance. A representative volume was mapped onto a two-dimensional square network where the porosity at a given location was obtained as the average of the binary data over the 50 slices. Thus, for each location, a porosity value varying from $0$ to $100\%$ was obtained.  A typical representative volume, shown in Fig.~\ref{fig:representative domain}(c), is fairly dense, having mean porosity lower than 5\%, and consists of  fine pores that are evenly distributed and extend across the height of the bone sample while being slightly inclined to the longitudinal direction. The two-dimensional porosity network, thus obtained from the CT scans, is an input for the RSNM model constructed in the following manner.

\begin{figure}[h!]
\centering
\includegraphics[width = 8cm]{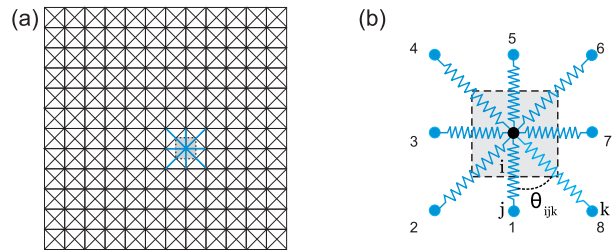}
\caption {\label{fig:schematic RSNM}(a) Schematic diagram of the spring network model. (b) Unit cell of the network showing the linear springs attached to a site. Also shown is an example of an angle $\theta_{ijk}$  whose distortion is resisted by a bending spring.}
\end{figure}

A representative volume is modeled using a $150 \times 150$ size square lattice where the nearest and next-nearest neighbor pairs of particles are connected by linear springs (see Fig.~\ref{fig:schematic RSNM}). In addition to the energy due to extension of the springs, we associate a bending energy for any deviation of the angle between two adjacent linear springs [see $\theta_{ijk}$ shown in Fig.~\ref{fig:schematic RSNM}(b)] from the initial value of $\pi/4$. The potential energy of the network, $V$, has contributions from extension of springs, distortion of angle between springs and also a repulsive contact force modeled as Hertzian contact between particles:
\begin{equation}
V\!\!=\!\!\sum_{\langle ij\rangle} \frac{k_{ij}}{2}\delta r_{ij}^2 + \sum_{\langle ijk\rangle}\frac{c_{ijk}}{2}  \delta \theta_{ijk}^2 + \alpha \sum_{mn} (d-r_{mn})^{\frac{3}{2}} \Theta(d-r_{mn}),
\end{equation} 
where $\langle ij \rangle$ denotes those pairs of particles that are connected by linear springs, the spring constant being $k_{ij}$ and extension being $\delta r_{ij}$. $c_{ijk}$ is the spring constant that resists the distortion in angle, $\delta \theta_{ijk}$, between triad of particles $\langle ijk \rangle$. The contact between any two particles $m$ and $n$ is initiated by a Heaviside function, $\Theta(x)$, only when the inter particle distance, $r_{mn}$, is less than the particle diameter, $d$. The elastic contact force parameter $\alpha$ is a material constant~\citep{landau1986}. If the springs constants of horizontal/vertical, diagonal and torsional springs  are $2 k$, $k$ and $c$ respectively, then the system is isotropic with elastic modulus, $E$, and Poisson's ratio, $\nu$, given by~\citep{monette}
\begin{equation}
E = \frac{8k(k+ca^{-2})}{3k+ca^{-2}};\quad
\nu = \frac{k-ca^{-2}}{3k+ca^{-2}},
\label{eq:modulus}
\end{equation}
where $a$ is the lattice spacing. 
We account for the elastic compliance of the testing machine by connecting the sites in the top and bottom rows of the spring network to springs whose compliance match with that of the machine used in experiments.
In the simulations, the displacements are applied incrementally from the top. For every increment, the system is equilibrated to its minimum energy configuration by numerically integrating the equations of motion in terms of the position vectors, $r_i$:
\begin{equation}
\frac{d^2 {\bf r}_i}{dt^2}=-\nabla_{{\bf r}_i} V- \gamma \frac{d {\bf r}_i}{dt},
\end{equation}
using velocity-verlet  algorithm~\cite{verlet}. The damping coefficient $\gamma$ dissipates energy and brings the system to its minimum energy configuration~\cite{ray}.
For each increment in the applied downward displacement, after equilibration, if any spring is stretched beyond its corresponding threshold strain $\epsilon_f$, it and the bending springs associated with it are broken. The system is re-equilibrated till no further breakage takes place within the increment.

To incorporate the experimentally obtained porosity network into the model, the spring constants and breaking strain thresholds are assigned in accordance with the local porosity. To find the porosity dependent material behavior, first the effective behavior of a porous network that has homogeneous matrix properties is evaluated. The response is evaluated under tension as the bone samples experience tensile transverse strains that lead to splitting under longitudinal compression.  A RSNM of given porosity, $P$, is constructed by removing voids [of size $50~\mu m$ for $0\leq P \leq 5\%$ and $100~\mu m$ for $5\%\leq P \leq 20\%$] at random from a homogenous matrix. From the macroscopic response of such porous networks, $E(P)$, $\nu(P)$ and $\epsilon_f(P)$ are obtained. The scaling of these curves depends on the values of the spring constants and breaking thresholds of the homogeneous matrix. These values are chosen such that at $4\%$ porosity, the known experimental average macroscopic response of bone is reproduced~\citep{li2013}. Next, for simulation of fracture in bone samples, the individual spring characteristics of the RSNM are assigned on the basis of the local porosity data from experiments using the corresponding calibrated  $E(P)$, $\nu(P)$ and $\epsilon_f(P)$ in Eq.(2) to achieve a porosity dependent RSNM.

To simulate the fracture process under compression loads, in a previous work, we developed a random spring network model where the matrix properties were taken to be independent of bone and the failure was predicted accounting only the structure of the porosity network~\citep{jrsoc}. While the porosity based RSNM  was shown to capture the characteristic features of the quantitative macroscopic response as well as the qualitative failure paths during the fracture process, quantitative correlations were poor. To improve the predictive capability of the model, here, we evaluate the use of  matrix properties that are specific to a bone. For validation, we use experimental data~\citep{jrsoc} on a total of six samples that were harvested from three different bovine femurs and referred to in the remainder of the paper as sample I and II from bovine-1, sample III and IV from bovine-2, and sample V and VI from bovine-3.

\section{Results and discussion}
Initially, a parametric study is performed to evaluate the sensitivity of the predictions to model parameters -- elastic modulus and failure strain. In the study, the matrix properties of the homogeneous porous network are varied by $\pm 10\%$ of the bone independent parameters used in \citet{jrsoc} that are  considered here as base values. 
\begin{figure}[h!]
\centering
\includegraphics[width = \columnwidth]{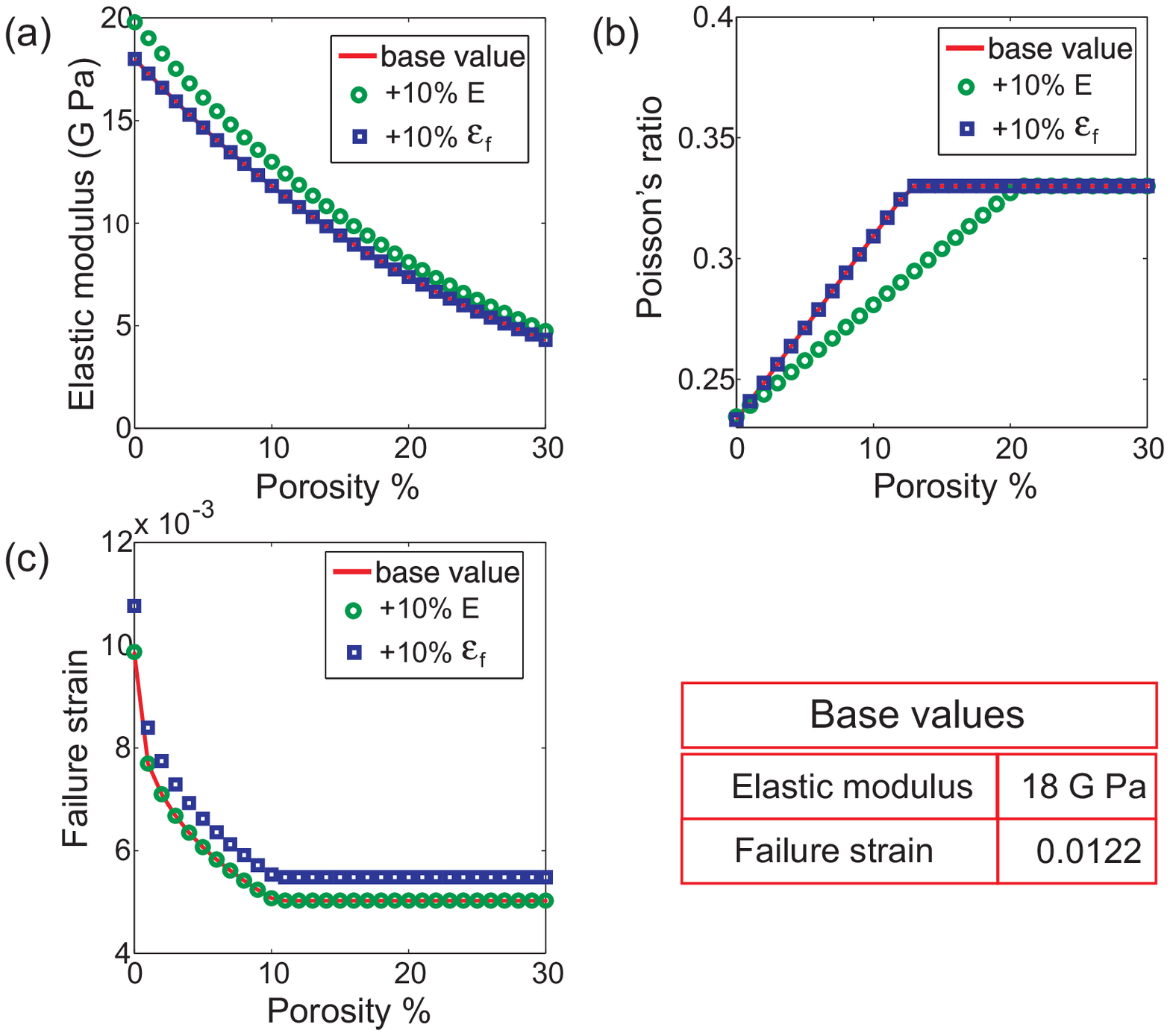}
\caption {\label{fig:parameters} Variation of (a) elastic modulus (b) Poisson's ratio and (c) failure strain with porosity for different modulus and fracture strain of the matrix.}
\end{figure}
Based on each combination of matrix properties, first the porosity dependent elastic modulus, Poisson's ratio and failure strain strain, as shown in Fig.~\ref{fig:parameters}, are evaluated from a porous network, as described in Sec II, for a range of overall porosity. For the cases when there is 10\% increase in $E$ and 10\% $\epsilon_f$ from the base value, as expected stiffer matrix results in stiffer macroscopic response and affects only the effective elastic behavior while change in the failure threshold correspondingly has effect only on the effective limiting strain and not on the elastic behavior. We note that, for thermodynamic stability, the spring constants $k$ and $c$ must be positive in Eq. (2), which bounds the Poisson ratio from above by 1/3.

\subsection{Effect on failure paths} 
\begin{figure}[htbp]
\centering
\includegraphics[width = \columnwidth]{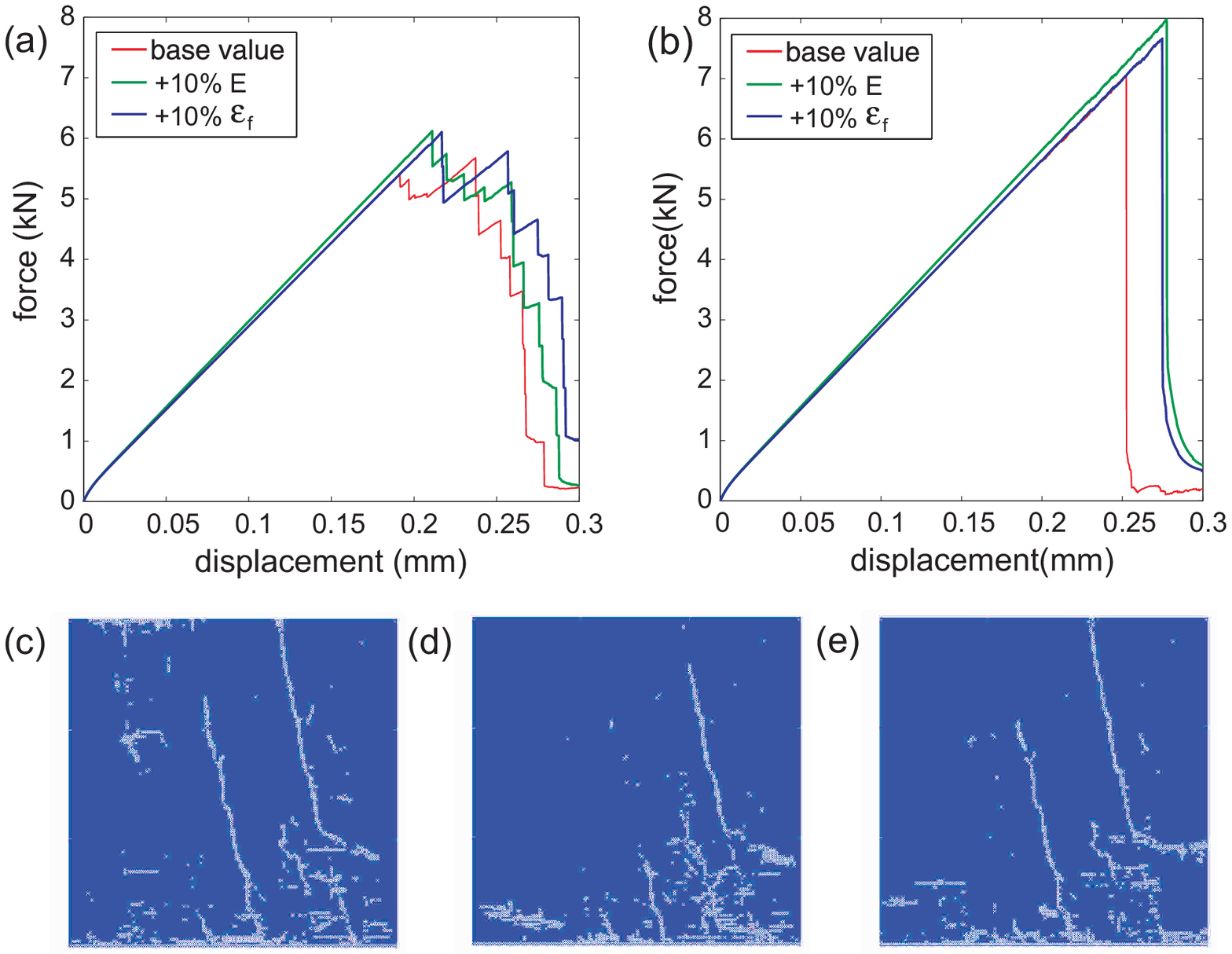}
\caption {\label{fig:failure paths} (a) Macroscopic response of sample III and (b) for a typical representative domain. The corresponding failure paths at the first load drop with (c) base values (d) 10\% $E$ and (e) 10\% failure strain as input parameters.}
\end{figure}
Using a porosity dependent RSNM, the effect of the model parameters: elastic modulus and failure strain on the macroscopic response is shown for a typical sample (sample III) in Fig.~\ref{fig:failure paths}. For clarity only the predictions based on the base value and 10\% independent increase in each of the model parameters is shown. The increase in $E$ as well as $\epsilon_f$ is shown to increase the overall load bearing capacity, obtained from the macroscopic response of the sample that has been averaged over the 10 representative domains, by approximately $8\%$. Most of the individual sets exhibit a similar increase in load bearing capacity as evident in Fig.~\ref{fig:failure paths}(b) that shows the macroscopic response of a typical representative domain. The corresponding failure paths are presented in Figs.~\ref{fig:failure paths}(c)-(e). The first bonds to fail for any combination of parameters are from the regions of relatively higher porosity ($10-20\%$) and form the primary failure paths. Increase in the failure threshold results in  minimal changes in the failure path as it involves no change in the load distribution and thus, the most critical zones remain unchanged. However, increase in $E$ could introduce differences in load transfer paths and thereby leads to some variations in the splitting paths. On rare occasions (one representative domain out of the 10 considered here), the damage at the critical defect localizes in a manner leading to the formation of a large single crack resulting in splitting fracture at lower strains. 

\subsection{Effect on macroscopic response} 
\begin{figure}[h!]
\centering
\includegraphics[width = 6.2cm]{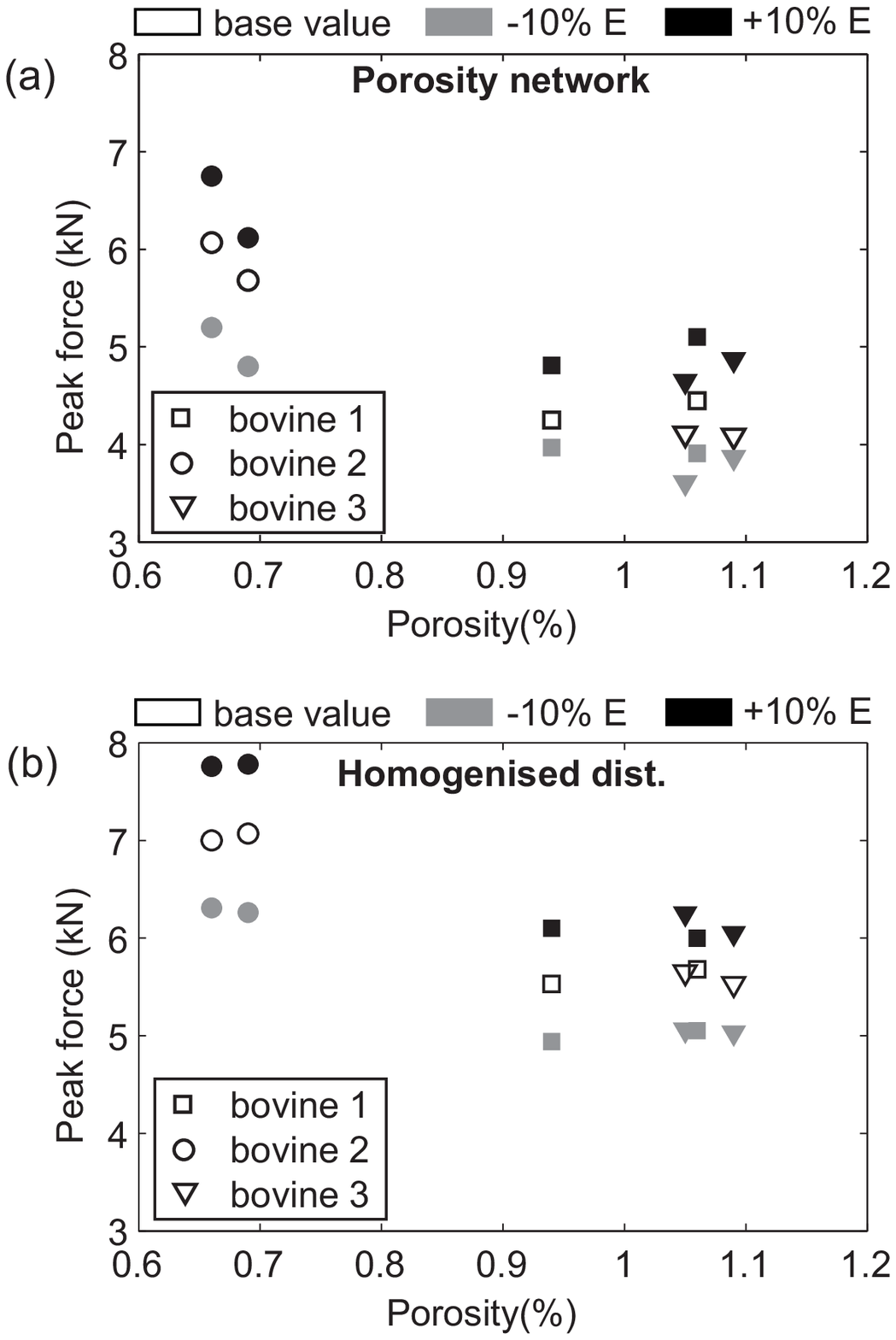}
\caption {\label{fig: elastic modulus}Variation of load bearing capacity with mean porosity levels and $\pm 10\%$ variation in elastic modulus, $E$, accounting (a) for  porosity network and (b) for homogenized distribution of porosity.}
\end{figure}

To develop a comprehensive understanding of the effect of model parameters on the macroscopic response under compression, simulations  were performed for all the 6 samples, using the structure of their respective porosity networks. Figure ~\ref{fig: elastic modulus}(a) shows the predicted load bearing capacity for elastic modulus ranging between $\pm 10\%$ of the base value as a function of the overall porosity of the sample. Also, for each sample a corresponding homogenized network is developed that has the equivalent porosity that is randomly distributed in the domain. The data generated thus is presented in Fig. ~\ref{fig: elastic modulus}(b). Irrespective of the inherent differences in the structure of the porosity network between samples, the load bearing capacity as predicted by the model decreases with increasing overall porosity. Also, for the same porosities and model parameters, the load bearing capacity of the homogenized network is approximately $28\%$ higher than  the corresponding networked porosity.
\begin{figure}[h!]
\centering
\includegraphics[width = 6.2cm]{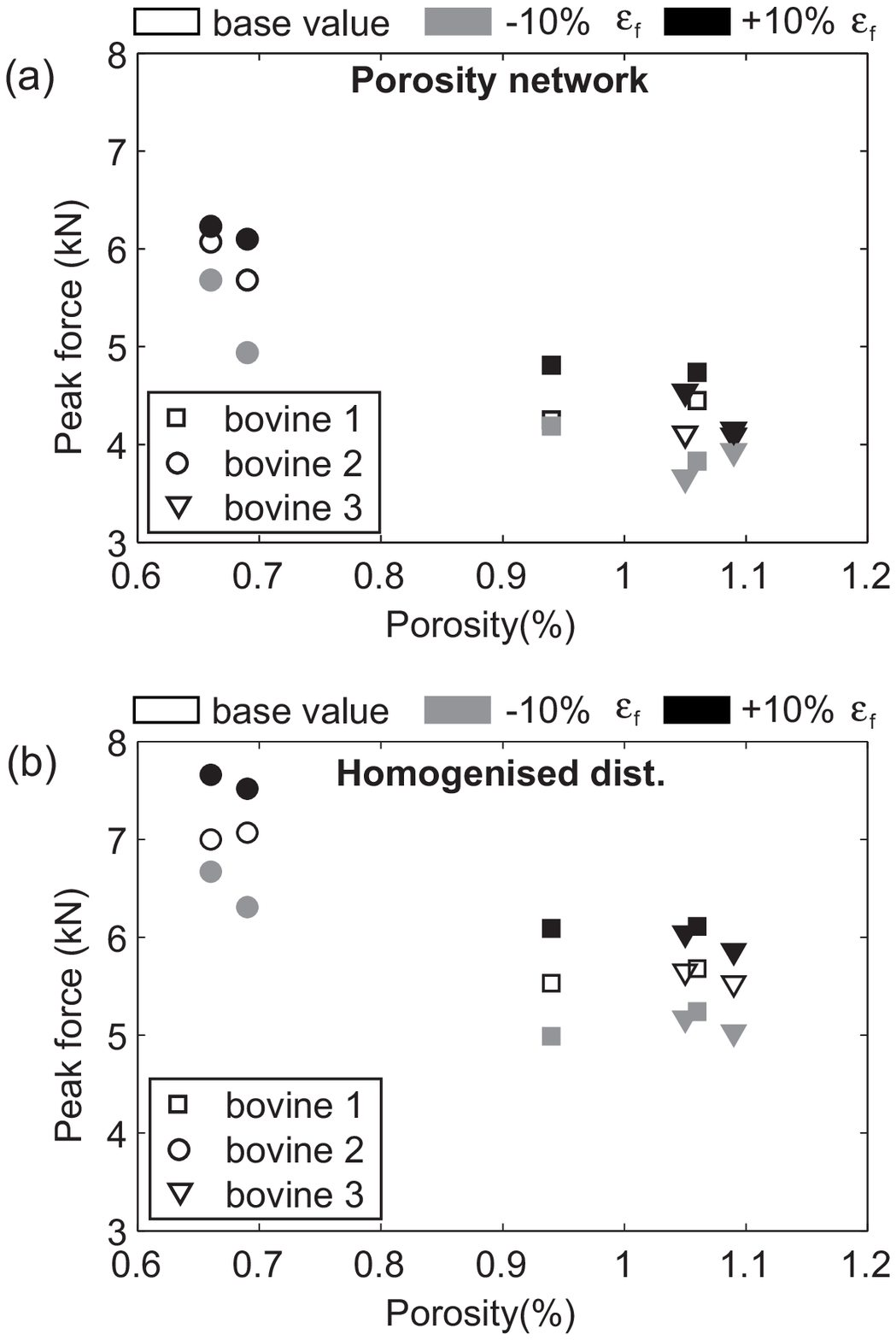}
\caption {\label{fig: failure threshold}Variation of load bearing capacity with mean porosity levels and $\pm 10\%$ variation in failure strain, $\epsilon_f$ accounting (a) for  porosity network and (b) for homogenized distribution of porosity.}
\end{figure}

Varying the elastic modulus from -10\% to +10\% of base value scales the predicted peak force, on an average, by $27\%$ as shown in Fig.~\ref{fig: elastic modulus}(a). Similar variations of the failure strain have relatively lower influence on the load bearing capacity -- on an average by $17\%$ as shown in Fig.~\ref{fig: failure threshold}(a). This significant increase can be attributed to the changes in the load transfer paths that may occur as a result of increase in elastic modulus and thereby resulting in differences in crack paths as discussed earlier. Homogenization of porosity network results in an overall increase in predictions of peak force values as shown in Figs.~\ref{fig: elastic modulus}(b) and \ref{fig: failure threshold}(b). As expected, the increase in the load bearing capacity of a homogeneously porous network is more comparable for the increase in either parameters: $22\%$ for increase in $E$ and $18\%$ for increase in $\epsilon_f$.

\subsection{Use of bone-specific properties}
\begin{figure}[h!]
\centering
\includegraphics[width = 7cm]{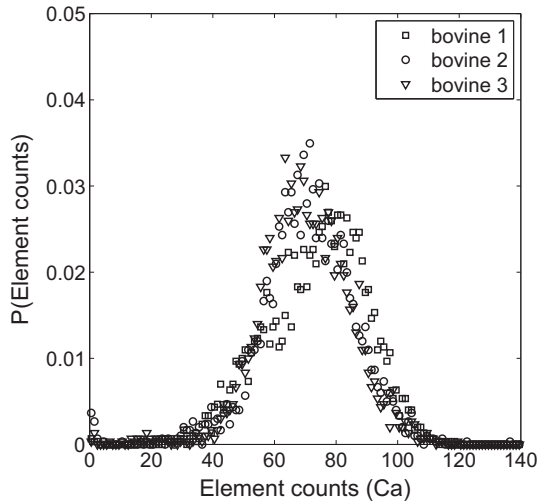}
\caption {\label{fig: histogram}Probability density of element counts from EDS line scans of bone samples.}
\end{figure}
To determine the appropriate bone-specific matrix properties for each bone, we first examine the elastic behavior experimentally. The elastic properties of cortical bone are known to be influenced by the mineral density~\citep{schaffler, currey_88} which are typically characterized by different techniques including electron micro-probe techniques such as energy dispersive spectroscopy (EDS)~\citep{aakesson,bloebaum}. Here, we estimate the differences in the elastic behavior of the bone matrix between bones by performing EDS line scans using Quanta-200 FEI scanning electron microscope. The sample measurements are grouped on the basis of the bovine femurs from which they are harvested. The probability density of calcium counts obtained for each bone, shown in Fig.~\ref{fig: histogram}, are bell shaped curves with significant overlap. The expectation values for the counts for the bones are with in $4\%$ of the overall average count which is possibly because the bones are from healthy animals of similar age. Thus, for further analysis the elastic behavior of all the bones is taken to be the base value. 

To incorporate the differences in the fracture behavior of bone matrix, the threshold failure strain for each bone was estimated iteratively till it compared well with the experimental data and is presented in Table~\ref{table1}.  

\begin{table}[h!]
\begin{ruledtabular}
\caption{\label{table1}Bone specific input parameters}
\begin{tabular}{c c c}
 Bone & Sample no.  &Parameter for simulations\\ \hline \hline
\multirow{2}{*}{bovine - 1} & I & \multirow{2}{*}{$E$ and $1.13\times \epsilon_f$}\\
                            & II & \\ \hline
\multirow{2}{*}{bovine - 2} & III & \multirow{2}{*}{$E$ and $0.9 \times \epsilon_f$}\\
                            & IV & \\ \hline
\multirow{2}{*}{bovine - 3} & V & \multirow{2}{*}{$E$ and $1.68\times \epsilon_f$}\\ 
                            & VI &   \\ \hline                                                
\multicolumn{3}{c}{* Base values : $E = 18~GPa$ and $\epsilon_f = 0.0122$}\\ 
\end{tabular}
\end{ruledtabular}
\end{table}

\begin{figure*}[t]
\centering
\includegraphics[width = 13.5cm]{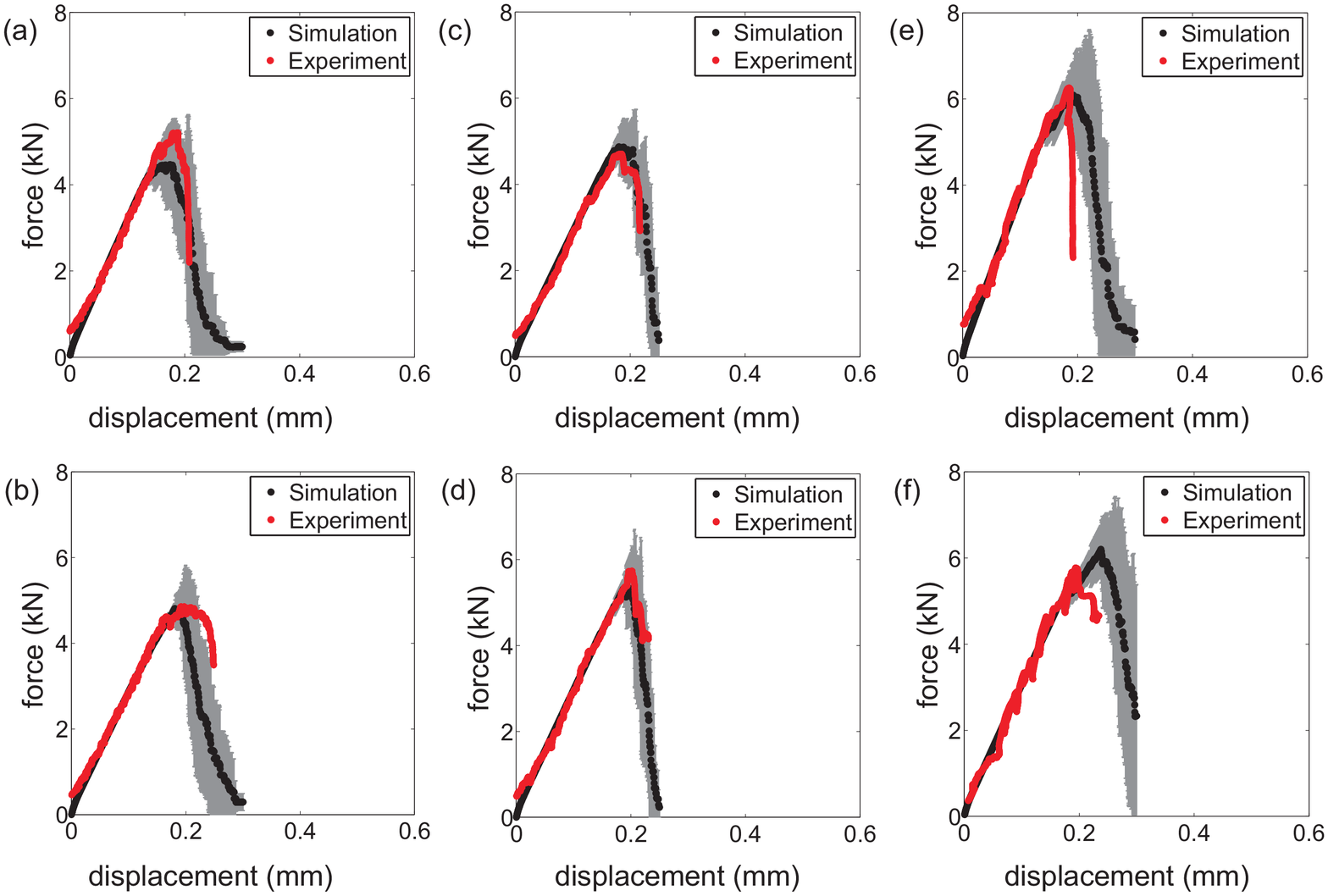}
\caption {\label{fig:fd plot} Comparison of macroscopic response from simulations and experiments of bovine samples, where (a) is for sample I, (b) is for sample II and so on.}
\end{figure*}

For a given sample, 5 random realizations for each of the 10 representative networks were performed. The strain threshold for each bond was taken from a Gaussian distributions with mean as in Table~\ref{table1} and 5\% standard deviation to account for heterogeneity at length scales much smaller than the lattice parameter. The macroscopic response, averaged over all the 50 simulations per sample are compared with the experimental data in Fig.~\ref{fig:fd plot}. The characteristic features of the macroscopic response such as the initial linear elasticity, the maximum load bearing capacity, multiple smaller events prior to final failure etc. are well reproduced. It is remarkable to note that for each bone, of the two samples tested, same model parameters are effective in predicting response that compares closely with experimental data for both the samples.

\begin{figure}[h!]
\centering
\includegraphics[width = 7.5cm]{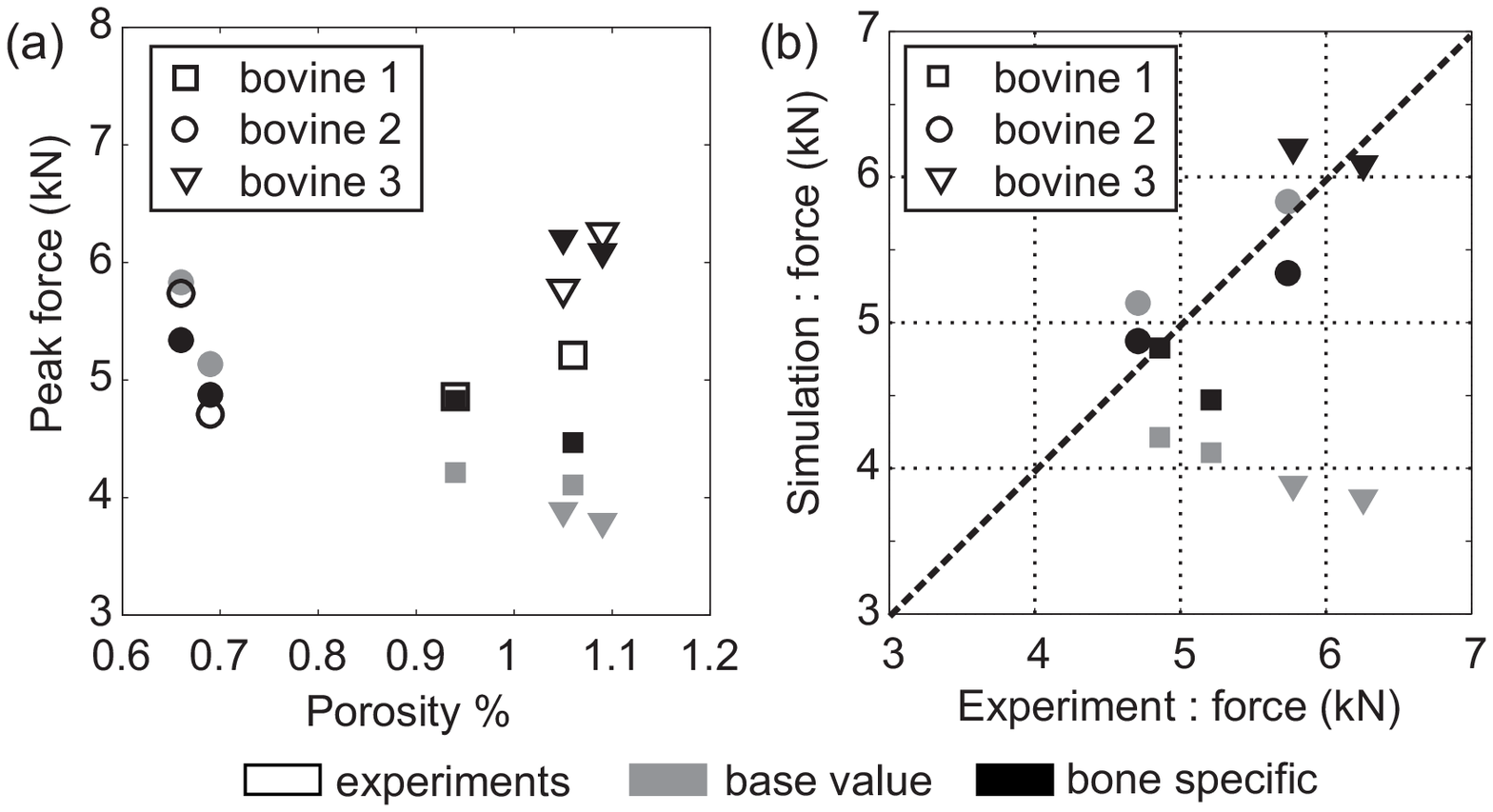}
\caption {\label{fig: bone_specific}Load bearing capacity from experiments and simulations: (a) variation with porosity and (b) correlation between experiment and simulations.}
\end{figure}

A comparative summary of the maximum load taken by any sample as per simulation based on bone specific properties, base values and as from experimental data is presented in Fig.~\ref{fig: bone_specific}(a). Maximum loads as predicted using base values for all samples, as seen earlier, tend to have a monotonic decrease with increasing porosity which is in contradiction to the experimental data where there is no such consistent trend with respect to porosity.  While the samples from bovine 2, bovine 1 and bovine 3 appear to be in order of increasing porosity, the load taken by them has no direct porosity dependence, in fact the highest load is taken by the most porous of the samples. However, when the bone specific properties are incorporated in the simulations a significant improvement in predictions are seen for both samples of each bone as illustrated in Fig.~\ref{fig: bone_specific}(b).
The concordance correlation coefficients from the simulations with bone-specific parameters~(0.79) are also significant as compared to base values~(-0.12). The differences in load bearing capacity between samples within a bone that are extracted from the same anatomical site can be attributed primarily to the inherent differences in the porosity network. However,  between samples from different bones with apparently similar elastic behavior, as per the model of the present study, the fracture behavior prediction requires 60-70\% difference in matrix failure strain threshold. These inferences require validation through detailed characterization of failure properties at smaller length scales.

\subsection{Avalanche size distribution}
One of the quantities that has been used in the literature to characterize fracture in heterogeneous media is the avalanche size distribution which measures the incremental response of the system to incremental increases in the external loading. In experiments, the response is measured through energy of acoustic emissions, $E$,  that occur during the fracture process. In simulations, it is measured by the number of events (broken springs), $s$, that occur per increment of strain. The probability distributions for these quantities are known to be power laws, implying the absence of a typical avalanche size that is independent of system size. The exponents characterizing the acoustic emission and avalanche size distribution may be related to each other. Let $P_E(E)$ and $P_s(s)$ denote the respective distributions. Asymptotically, they behave as $P_E(E) \sim E^{-\tau_E}$ and $P_s(s) \sim s^{-\tau_s}$. However, it is known that $E \propto s^2$~\citep{ray}. Using the relation $P_E(E) dE = P_s(s) ds$, arising from the conservation of probability, it is straightforward to obtain $\tau_E= (1+\tau_s)/2$. Remarkably, these exponents are quite universal and independent of details of material or modeling. For example, $\tau_E$ is known to be $1.3-2.0$ for many brittle materials like synthetic plaster~\citep{petri1994}, wood~\citep{garcimartin1997}, fiberglass~\citep{guarino2002}, cellular glass~\citep{maes1998} and rocks~\citep{pradhan2015}. For bone under compressive loading, it has recently been reported for porcine bone that $\tau_E = 1.3-1.7$~\citep{baro16pig}. We now measure $\tau_s$ from our simulations, and ask whether and how the exponent $\tau_s$ depends on the choice of bone-dependent material properties, or the porosity network.
\begin{figure}[h!]
\centering
\includegraphics[width = 7cm]{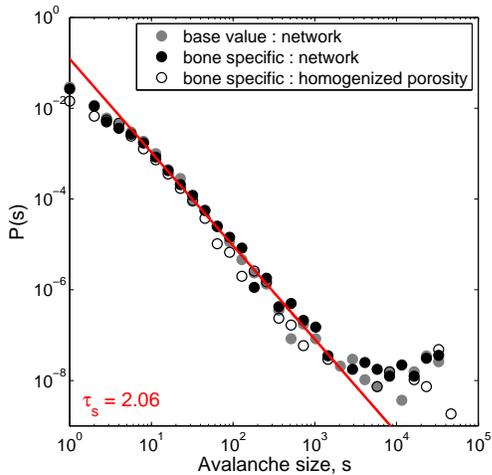}
\caption {\label{fig: avalanche size} Avalanche distribution $P(s)$ for sample III from simulations.}
\end{figure}

The avalanche size distribution for a typical sample is shown in Fig.~\ref{fig: avalanche size}. For each of the representative domains, for all 5 realizations, avalanche sizes are obtained for all increments in displacement. As can be seen from Fig.~\ref{fig: avalanche size}, the distribution is insensitive to whether the bone-specific parameters or base values are used. This is understandable as the bone specific parameters involve only a change in $\epsilon_f$, and this does not appreciably change the load distribution in the network during compression. For the sample shown in Fig.~\ref{fig: avalanche size}, we find $\tau_s \approx 2.06$ corresponding to $\tau_E=1.53$. For the other samples, the exponents that we obtain are shown in Table~\ref{table2}. The exponent values for $\tau_E$ lie in the range $1.35-1.5$, consistent with the values of $1.3-1.7$ obtained experimentally for porcine bone~\citep{baro16pig}. We also do not detect any significant correlation of the values of the exponent with the porosity of the sample. In Fig.~\ref{fig: avalanche size}, we also present the avalanche size distribution associated with an equivalent homogenized porous network. Surprisingly, the avalanche exponent does not change noticeably. However, the distribution for small avalanche sizes is different, reflecting the contrast in the network at small scales.
\begin{table}[h!]
\begin{ruledtabular}
\caption{\label{table2} Power law exponents, $\tau_s$ and $\tau_E$ for all samples.}
\begin{tabular}{c c c c}
 Bone & Sample no.  & $\tau_s$ & $\tau_E$\\ \hline \hline
\multirow{2}{*}{bovine - 1} & I &1.93  &1.47\\ 
                            & II & 1.80&1.40\\ \hline
\multirow{2}{*}{bovine - 2} & III &2.06 &1.53\\
                            & IV &1.93 &1.47\\ \hline
\multirow{2}{*}{bovine - 3} & V & 1.69 &1.35\\ 
                            & VI &1.93  &1.47 \\                                               
\end{tabular}
\end{ruledtabular}
\end{table}

\section{SUMMARY AND CONCLUSIONS}
Complex fracture processes in porous brittle materials are controlled by several factors. Using a porosity dependent RSNM, in the present study, we examine the individual roles of overall mean porosity, the networking of the porosity, the elastic behavior of non-porous matrix and the failure behavior of the non-porous matrix on the compressive strength of dry plexiform bone. As per the model, increasing mean porosity results in reduced compressive strength. Porosity network structure, as seen in plexiform bone, reduces the  compressive strength further by upto $30\%$. While the initiation of the fracture process is typically at regions of highest porosity, the crack pathways are predominantly controlled by the connectivity of the porosity network. Among the multiple competing porosity pathways, the damage localizes in only a few as driven by the load distribution which in turn is influenced  by the elastic properties of the model matrix and not by the failure strain threshold of the matrix. However, the load bearing capacity increases with increase in either elastic modulus or failure strain threshold of the matrix.

To validate the porosity-based RSNM model with available experimental data, bone-specific properties were applied. Of the six samples (two each from three different bones), the elastic properties were found to differ minimally as reflected in the mineral composition determined by EDS. Assuming elastic similarity, bone-specific failure strain thresholds for the ideal matrix were estimated to be within $60\%$ of each other. It is to be noted that the compressive strengths differ between the two samples of the same bone and yet, identical bone-specific model parameters achieve excellent concordance correlation between experiment and simulations for both samples of each bone.  To provide an accurate input for the model, it would therefore be of importance to develop an experimental scheme that characterizes the strain threshold of the matrix at small length scales.

We find that the avalanche size exponents are independent of the bone-dependent parameters and in the range of the experimental values obtained for porcine bone~\citep{baro16pig}.  Surprisingly, the exponent is also independent of the structure of the porosity network as the avalanche size distribution exponent for the equivalent homogenized network has similar value. This may be the reason why the avalanche exponent is universal and has similar numerical value for many different kinds of material. At the same time, the exponents obtained in this paper ($1.7-2. 05$) are significantly different from that for homogeneous RSNM under tension ($\sim 2.5$)~\citep{ray}. Whether this difference is due to the difference in the nature of loading, or it is because of the strong correlation between the values for the spring constant and its strain threshold, as in the current study, remains to be answered.

It would be interesting to study the predictions of the model for Haversian bone which has a distinctly different microstructure from plexiform bone. Unlike the layered plexiform bone, both porosity network characteristics and material constitution are different that may influence the load bearing capacity. A similar analysis for Haversian bone  is part of ongoing studies.

\begin{acknowledgements}
We thank Prof. Krishnan Balasubramanian, Centre for Non-Destructive Evaluation and Prof. V. S. Sarma, Scanning Electron Microscopy Lab, IIT Madras for providing access to the experimental facilities.
\end{acknowledgements}


\begin{thebibliography}{58}%
\makeatletter
\providecommand \@ifxundefined [1]{%
 \@ifx{#1\undefined}
}%
\providecommand \@ifnum [1]{%
 \ifnum #1\expandafter \@firstoftwo
 \else \expandafter \@secondoftwo
 \fi
}%
\providecommand \@ifx [1]{%
 \ifx #1\expandafter \@firstoftwo
 \else \expandafter \@secondoftwo
 \fi
}%
\providecommand \natexlab [1]{#1}%
\providecommand \enquote  [1]{``#1''}%
\providecommand \bibnamefont  [1]{#1}%
\providecommand \bibfnamefont [1]{#1}%
\providecommand \citenamefont [1]{#1}%
\providecommand \href@noop [0]{\@secondoftwo}%
\providecommand \href [0]{\begingroup \@sanitize@url \@href}%
\providecommand \@href[1]{\@@startlink{#1}\@@href}%
\providecommand \@@href[1]{\endgroup#1\@@endlink}%
\providecommand \@sanitize@url [0]{\catcode `\\12\catcode `\$12\catcode
  `\&12\catcode `\#12\catcode `\^12\catcode `\_12\catcode `\%12\relax}%
\providecommand \@@startlink[1]{}%
\providecommand \@@endlink[0]{}%
\providecommand \url  [0]{\begingroup\@sanitize@url \@url }%
\providecommand \@url [1]{\endgroup\@href {#1}{\urlprefix }}%
\providecommand \urlprefix  [0]{URL }%
\providecommand \Eprint [0]{\href }%
\providecommand \doibase [0]{http://dx.doi.org/}%
\providecommand \selectlanguage [0]{\@gobble}%
\providecommand \bibinfo  [0]{\@secondoftwo}%
\providecommand \bibfield  [0]{\@secondoftwo}%
\providecommand \translation [1]{[#1]}%
\providecommand \BibitemOpen [0]{}%
\providecommand \bibitemStop [0]{}%
\providecommand \bibitemNoStop [0]{.\EOS\space}%
\providecommand \EOS [0]{\spacefactor3000\relax}%
\providecommand \BibitemShut  [1]{\csname bibitem#1\endcsname}%
\let\auto@bib@innerbib\@empty
\bibitem [{\citenamefont {Pearce}\ \emph {et~al.}(2007)\citenamefont {Pearce},
  \citenamefont {Richards}, \citenamefont {Milz}, \citenamefont {Schneider},
  \citenamefont {Pearce} \emph {et~al.}}]{pearce}%
  \BibitemOpen
  \bibfield  {author} {\bibinfo {author} {\bibfnamefont {A.}~\bibnamefont
  {Pearce}}, \bibinfo {author} {\bibfnamefont {R.}~\bibnamefont {Richards}},
  \bibinfo {author} {\bibfnamefont {S.}~\bibnamefont {Milz}}, \bibinfo {author}
  {\bibfnamefont {E.}~\bibnamefont {Schneider}}, \bibinfo {author}
  {\bibfnamefont {S.}~\bibnamefont {Pearce}},  \emph {et~al.},\ }\href@noop {}
  {\bibfield  {journal} {\bibinfo  {journal} {Eur. Cell Mater.}\ }\textbf
  {\bibinfo {volume} {13}},\ \bibinfo {pages} {1} (\bibinfo {year}
  {2007})}\BibitemShut {NoStop}%
\bibitem [{\citenamefont {Conward}\ and\ \citenamefont
  {Samuel}(2016)}]{conward2016}%
  \BibitemOpen
  \bibfield  {author} {\bibinfo {author} {\bibfnamefont {M.}~\bibnamefont
  {Conward}}\ and\ \bibinfo {author} {\bibfnamefont {J.}~\bibnamefont
  {Samuel}},\ }\href@noop {} {\bibfield  {journal} {\bibinfo  {journal} {J.
  Mech. Behav. Biomed. Mater.}\ }\textbf {\bibinfo {volume} {60}},\ \bibinfo
  {pages} {525} (\bibinfo {year} {2016})}\BibitemShut {NoStop}%
\bibitem [{\citenamefont {Hing}(2005)}]{hing}%
  \BibitemOpen
  \bibfield  {author} {\bibinfo {author} {\bibfnamefont {K.~A.}\ \bibnamefont
  {Hing}},\ }\href@noop {} {\bibfield  {journal} {\bibinfo  {journal} {Int. J.
  Appl. Ceram. Tec.}\ }\textbf {\bibinfo {volume} {2}},\ \bibinfo {pages} {184}
  (\bibinfo {year} {2005})}\BibitemShut {NoStop}%
\bibitem [{\citenamefont {Bansiddhi}\ \emph {et~al.}(2008)\citenamefont
  {Bansiddhi}, \citenamefont {Sargeant}, \citenamefont {Stupp},\ and\
  \citenamefont {Dunand}}]{bansiddhi}%
  \BibitemOpen
  \bibfield  {author} {\bibinfo {author} {\bibfnamefont {A.}~\bibnamefont
  {Bansiddhi}}, \bibinfo {author} {\bibfnamefont {T.}~\bibnamefont {Sargeant}},
  \bibinfo {author} {\bibfnamefont {S.~I.}\ \bibnamefont {Stupp}}, \ and\
  \bibinfo {author} {\bibfnamefont {D.}~\bibnamefont {Dunand}},\ }\href@noop {}
  {\bibfield  {journal} {\bibinfo  {journal} {Acta Biomater.}\ }\textbf
  {\bibinfo {volume} {4}},\ \bibinfo {pages} {773} (\bibinfo {year}
  {2008})}\BibitemShut {NoStop}%
\bibitem [{\citenamefont {Libonati}\ and\ \citenamefont
  {Vergani}(2016)}]{libonati2016}%
  \BibitemOpen
  \bibfield  {author} {\bibinfo {author} {\bibfnamefont {F.}~\bibnamefont
  {Libonati}}\ and\ \bibinfo {author} {\bibfnamefont {L.}~\bibnamefont
  {Vergani}},\ }\href@noop {} {\bibfield  {journal} {\bibinfo  {journal}
  {Compos. Struct.}\ }\textbf {\bibinfo {volume} {139}},\ \bibinfo {pages}
  {188} (\bibinfo {year} {2016})}\BibitemShut {NoStop}%
\bibitem [{\citenamefont {Chen}\ \emph {et~al.}(2005)\citenamefont {Chen},
  \citenamefont {Li}, \citenamefont {Lu}, \citenamefont {Tang}, \citenamefont
  {Sun},\ and\ \citenamefont {Xu}}]{chen}%
  \BibitemOpen
  \bibfield  {author} {\bibinfo {author} {\bibfnamefont {Z.}~\bibnamefont
  {Chen}}, \bibinfo {author} {\bibfnamefont {D.}~\bibnamefont {Li}}, \bibinfo
  {author} {\bibfnamefont {B.}~\bibnamefont {Lu}}, \bibinfo {author}
  {\bibfnamefont {Y.}~\bibnamefont {Tang}}, \bibinfo {author} {\bibfnamefont
  {M.}~\bibnamefont {Sun}}, \ and\ \bibinfo {author} {\bibfnamefont
  {S.}~\bibnamefont {Xu}},\ }\href@noop {} {\bibfield  {journal} {\bibinfo
  {journal} {Scripta Mater.}\ }\textbf {\bibinfo {volume} {52}},\ \bibinfo
  {pages} {157} (\bibinfo {year} {2005})}\BibitemShut {NoStop}%
\bibitem [{\citenamefont {Liu}\ \emph {et~al.}(2013)\citenamefont {Liu},
  \citenamefont {Rahaman},\ and\ \citenamefont {Fu}}]{liu}%
  \BibitemOpen
  \bibfield  {author} {\bibinfo {author} {\bibfnamefont {X.}~\bibnamefont
  {Liu}}, \bibinfo {author} {\bibfnamefont {M.~N.}\ \bibnamefont {Rahaman}}, \
  and\ \bibinfo {author} {\bibfnamefont {Q.}~\bibnamefont {Fu}},\ }\href@noop
  {} {\bibfield  {journal} {\bibinfo  {journal} {Acta Biomater.}\ }\textbf
  {\bibinfo {volume} {9}},\ \bibinfo {pages} {4889} (\bibinfo {year}
  {2013})}\BibitemShut {NoStop}%
\bibitem [{\citenamefont {Keenan}\ \emph {et~al.}(2017)\citenamefont {Keenan},
  \citenamefont {Mears},\ and\ \citenamefont {Skedros}}]{keenan2017}%
  \BibitemOpen
  \bibfield  {author} {\bibinfo {author} {\bibfnamefont {K.~E.}\ \bibnamefont
  {Keenan}}, \bibinfo {author} {\bibfnamefont {C.~S.}\ \bibnamefont {Mears}}, \
  and\ \bibinfo {author} {\bibfnamefont {J.~G.}\ \bibnamefont {Skedros}},\
  }\href@noop {} {\bibfield  {journal} {\bibinfo  {journal} {Am. J. Phys.
  Anthropol.}\ }\textbf {\bibinfo {volume} {162}},\ \bibinfo {pages} {657}
  (\bibinfo {year} {2017})}\BibitemShut {NoStop}%
\bibitem [{\citenamefont {Zedda}\ \emph {et~al.}()\citenamefont {Zedda},
  \citenamefont {Palombo}, \citenamefont {Brits}, \citenamefont {Carcupino},
  \citenamefont {Sath{\'e}}, \citenamefont {Cacchioli},\ and\ \citenamefont
  {Farina}}]{zeddadifferences}%
  \BibitemOpen
  \bibfield  {author} {\bibinfo {author} {\bibfnamefont {M.}~\bibnamefont
  {Zedda}}, \bibinfo {author} {\bibfnamefont {M.~R.}\ \bibnamefont {Palombo}},
  \bibinfo {author} {\bibfnamefont {D.}~\bibnamefont {Brits}}, \bibinfo
  {author} {\bibfnamefont {M.}~\bibnamefont {Carcupino}}, \bibinfo {author}
  {\bibfnamefont {V.}~\bibnamefont {Sath{\'e}}}, \bibinfo {author}
  {\bibfnamefont {A.}~\bibnamefont {Cacchioli}}, \ and\ \bibinfo {author}
  {\bibfnamefont {V.}~\bibnamefont {Farina}},\ }\href@noop {} {\bibinfo
  {journal} {Zoomorphology}\ ,\ \bibinfo {pages} {1}}\BibitemShut {NoStop}%
\bibitem [{\citenamefont {Mitchell}\ \emph {et~al.}(2017)\citenamefont
  {Mitchell}, \citenamefont {Legendre}, \citenamefont {Lef{\`e}vre},\ and\
  \citenamefont {Cubo}}]{mitchell2017}%
  \BibitemOpen
\bibfield  {journal} {  }\bibfield  {author} {\bibinfo {author} {\bibfnamefont
  {J.}~\bibnamefont {Mitchell}}, \bibinfo {author} {\bibfnamefont {L.~J.}\
  \bibnamefont {Legendre}}, \bibinfo {author} {\bibfnamefont {C.}~\bibnamefont
  {Lef{\`e}vre}}, \ and\ \bibinfo {author} {\bibfnamefont {J.}~\bibnamefont
  {Cubo}},\ }\href@noop {} {\bibfield  {journal} {\bibinfo  {journal}
  {Zoology}\ } (\bibinfo {year} {2017})}\BibitemShut {NoStop}%
\bibitem [{\citenamefont {Claes}\ \emph {et~al.}(1995)\citenamefont {Claes},
  \citenamefont {Wilke},\ and\ \citenamefont {Kiefer}}]{claes}%
  \BibitemOpen
  \bibfield  {author} {\bibinfo {author} {\bibfnamefont {L.~E.}\ \bibnamefont
  {Claes}}, \bibinfo {author} {\bibfnamefont {H.-J.}\ \bibnamefont {Wilke}}, \
  and\ \bibinfo {author} {\bibfnamefont {H.}~\bibnamefont {Kiefer}},\
  }\href@noop {} {\bibfield  {journal} {\bibinfo  {journal} {J. Biomech.}\
  }\textbf {\bibinfo {volume} {28}},\ \bibinfo {pages} {1377} (\bibinfo {year}
  {1995})}\BibitemShut {NoStop}%
\bibitem [{\citenamefont {Gocha}\ and\ \citenamefont
  {Agnew}(2015)}]{gocha2015}%
  \BibitemOpen
  \bibfield  {author} {\bibinfo {author} {\bibfnamefont {T.~P.}\ \bibnamefont
  {Gocha}}\ and\ \bibinfo {author} {\bibfnamefont {A.~M.}\ \bibnamefont
  {Agnew}},\ }\href@noop {} {\bibfield  {journal} {\bibinfo  {journal} {J.
  Anat.}\ } (\bibinfo {year} {2015})}\BibitemShut {NoStop}%
\bibitem [{\citenamefont {Meyers}\ and\ \citenamefont {Chawla}(2009)}]{meyers}%
  \BibitemOpen
  \bibfield  {author} {\bibinfo {author} {\bibfnamefont {M.~A.}\ \bibnamefont
  {Meyers}}\ and\ \bibinfo {author} {\bibfnamefont {K.~K.}\ \bibnamefont
  {Chawla}},\ }\href@noop {} {\emph {\bibinfo {title} {Mechanical behavior of
  materials}}},\ Vol.~\bibinfo {volume} {2}\ (\bibinfo  {publisher} {Cambridge
  Univ. Press Cambridge},\ \bibinfo {year} {2009})\BibitemShut {NoStop}%
\bibitem [{\citenamefont {Shojaei}\ \emph {et~al.}(2014)\citenamefont
  {Shojaei}, \citenamefont {Taleghani},\ and\ \citenamefont {Li}}]{shojaei}%
  \BibitemOpen
  \bibfield  {author} {\bibinfo {author} {\bibfnamefont {A.}~\bibnamefont
  {Shojaei}}, \bibinfo {author} {\bibfnamefont {A.~D.}\ \bibnamefont
  {Taleghani}}, \ and\ \bibinfo {author} {\bibfnamefont {G.}~\bibnamefont
  {Li}},\ }\href@noop {} {\bibfield  {journal} {\bibinfo  {journal} {Int. J.
  Plast.}\ }\textbf {\bibinfo {volume} {59}},\ \bibinfo {pages} {199} (\bibinfo
  {year} {2014})}\BibitemShut {NoStop}%
\bibitem [{\citenamefont {Reilly}\ and\ \citenamefont
  {Burstein}(1975)}]{reilly}%
  \BibitemOpen
  \bibfield  {author} {\bibinfo {author} {\bibfnamefont {D.~T.}\ \bibnamefont
  {Reilly}}\ and\ \bibinfo {author} {\bibfnamefont {A.~H.}\ \bibnamefont
  {Burstein}},\ }\href@noop {} {\bibfield  {journal} {\bibinfo  {journal} {J.
  Biomech.}\ }\textbf {\bibinfo {volume} {8}},\ \bibinfo {pages} {393}
  (\bibinfo {year} {1975})}\BibitemShut {NoStop}%
\bibitem [{\citenamefont {Lipson}\ and\ \citenamefont {Katz}(1984)}]{katz_1}%
  \BibitemOpen
  \bibfield  {author} {\bibinfo {author} {\bibfnamefont {S.~F.}\ \bibnamefont
  {Lipson}}\ and\ \bibinfo {author} {\bibfnamefont {J.~L.}\ \bibnamefont
  {Katz}},\ }\href@noop {} {\bibfield  {journal} {\bibinfo  {journal} {J.
  Biomech.}\ }\textbf {\bibinfo {volume} {17}},\ \bibinfo {pages} {231}
  (\bibinfo {year} {1984})}\BibitemShut {NoStop}%
\bibitem [{\citenamefont {Katz}\ \emph {et~al.}(1984)\citenamefont {Katz},
  \citenamefont {Yoon}, \citenamefont {Lipson}, \citenamefont {Maharidge},
  \citenamefont {Meunier},\ and\ \citenamefont {Christel}}]{katz_2}%
  \BibitemOpen
  \bibfield  {author} {\bibinfo {author} {\bibfnamefont {J.~L.}\ \bibnamefont
  {Katz}}, \bibinfo {author} {\bibfnamefont {H.~S.}\ \bibnamefont {Yoon}},
  \bibinfo {author} {\bibfnamefont {S.}~\bibnamefont {Lipson}}, \bibinfo
  {author} {\bibfnamefont {R.}~\bibnamefont {Maharidge}}, \bibinfo {author}
  {\bibfnamefont {A.}~\bibnamefont {Meunier}}, \ and\ \bibinfo {author}
  {\bibfnamefont {P.}~\bibnamefont {Christel}},\ }\href@noop {} {\bibfield
  {journal} {\bibinfo  {journal} {Calcif. Tissue Int.}\ }\textbf {\bibinfo
  {volume} {36}},\ \bibinfo {pages} {S31} (\bibinfo {year} {1984})}\BibitemShut
  {NoStop}%
\bibitem [{\citenamefont {Shahar}\ \emph {et~al.}(2007)\citenamefont {Shahar},
  \citenamefont {Zaslansky}, \citenamefont {Barak}, \citenamefont {Friesem},
  \citenamefont {Currey},\ and\ \citenamefont {Weiner}}]{shahar}%
  \BibitemOpen
  \bibfield  {author} {\bibinfo {author} {\bibfnamefont {R.}~\bibnamefont
  {Shahar}}, \bibinfo {author} {\bibfnamefont {P.}~\bibnamefont {Zaslansky}},
  \bibinfo {author} {\bibfnamefont {M.}~\bibnamefont {Barak}}, \bibinfo
  {author} {\bibfnamefont {A.}~\bibnamefont {Friesem}}, \bibinfo {author}
  {\bibfnamefont {J.}~\bibnamefont {Currey}}, \ and\ \bibinfo {author}
  {\bibfnamefont {S.}~\bibnamefont {Weiner}},\ }\href@noop {} {\bibfield
  {journal} {\bibinfo  {journal} {J. Biomech.}\ }\textbf {\bibinfo {volume}
  {40}},\ \bibinfo {pages} {252} (\bibinfo {year} {2007})}\BibitemShut
  {NoStop}%
\bibitem [{\citenamefont {Martin}\ and\ \citenamefont
  {Ishida}(1989)}]{martin_1}%
  \BibitemOpen
  \bibfield  {author} {\bibinfo {author} {\bibfnamefont {R.}~\bibnamefont
  {Martin}}\ and\ \bibinfo {author} {\bibfnamefont {J.}~\bibnamefont
  {Ishida}},\ }\href@noop {} {\bibfield  {journal} {\bibinfo  {journal} {J.
  Biomech.}\ }\textbf {\bibinfo {volume} {22}},\ \bibinfo {pages} {419}
  (\bibinfo {year} {1989})}\BibitemShut {NoStop}%
\bibitem [{\citenamefont {Martin}\ and\ \citenamefont
  {Boardman}(1993)}]{martin_2}%
  \BibitemOpen
  \bibfield  {author} {\bibinfo {author} {\bibfnamefont {R.}~\bibnamefont
  {Martin}}\ and\ \bibinfo {author} {\bibfnamefont {D.}~\bibnamefont
  {Boardman}},\ }\href@noop {} {\bibfield  {journal} {\bibinfo  {journal} {J.
  Biomech.}\ }\textbf {\bibinfo {volume} {26}},\ \bibinfo {pages} {1047}
  (\bibinfo {year} {1993})}\BibitemShut {NoStop}%
\bibitem [{\citenamefont {Norman}\ \emph {et~al.}(1995)\citenamefont {Norman},
  \citenamefont {Vashishth},\ and\ \citenamefont {Burr}}]{norman}%
  \BibitemOpen
  \bibfield  {author} {\bibinfo {author} {\bibfnamefont {T.~L.}\ \bibnamefont
  {Norman}}, \bibinfo {author} {\bibfnamefont {D.}~\bibnamefont {Vashishth}}, \
  and\ \bibinfo {author} {\bibfnamefont {D.~B.}\ \bibnamefont {Burr}},\
  }\href@noop {} {\bibfield  {journal} {\bibinfo  {journal} {J. Biomech.}\
  }\textbf {\bibinfo {volume} {28}},\ \bibinfo {pages} {309} (\bibinfo {year}
  {1995})}\BibitemShut {NoStop}%
\bibitem [{\citenamefont {Reilly}\ and\ \citenamefont
  {Currey}(1999)}]{currey_2}%
  \BibitemOpen
  \bibfield  {author} {\bibinfo {author} {\bibfnamefont {G.~C.}\ \bibnamefont
  {Reilly}}\ and\ \bibinfo {author} {\bibfnamefont {J.~D.}\ \bibnamefont
  {Currey}},\ }\href@noop {} {\bibfield  {journal} {\bibinfo  {journal} {J.
  Exp. Biol.}\ }\textbf {\bibinfo {volume} {202}},\ \bibinfo {pages} {543}
  (\bibinfo {year} {1999})}\BibitemShut {NoStop}%
\bibitem [{\citenamefont {Kim}\ \emph {et~al.}(2005)\citenamefont {Kim},
  \citenamefont {Niinomi}, \citenamefont {Akahori}, \citenamefont {Takeda},\
  and\ \citenamefont {Toda}}]{kim2005}%
  \BibitemOpen
  \bibfield  {author} {\bibinfo {author} {\bibfnamefont {J.~H.}\ \bibnamefont
  {Kim}}, \bibinfo {author} {\bibfnamefont {M.}~\bibnamefont {Niinomi}},
  \bibinfo {author} {\bibfnamefont {T.}~\bibnamefont {Akahori}}, \bibinfo
  {author} {\bibfnamefont {J.}~\bibnamefont {Takeda}}, \ and\ \bibinfo {author}
  {\bibfnamefont {H.}~\bibnamefont {Toda}},\ }\href@noop {} {\bibfield
  {journal} {\bibinfo  {journal} {JSME Int J., Ser. A}\ }\textbf {\bibinfo
  {volume} {48}},\ \bibinfo {pages} {472} (\bibinfo {year} {2005})}\BibitemShut
  {NoStop}%
\bibitem [{\citenamefont {Kim}\ \emph {et~al.}(2007)\citenamefont {Kim},
  \citenamefont {Niinomi}, \citenamefont {Akahori},\ and\ \citenamefont
  {Toda}}]{niinomi}%
  \BibitemOpen
  \bibfield  {author} {\bibinfo {author} {\bibfnamefont {J.}~\bibnamefont
  {Kim}}, \bibinfo {author} {\bibfnamefont {M.}~\bibnamefont {Niinomi}},
  \bibinfo {author} {\bibfnamefont {T.}~\bibnamefont {Akahori}}, \ and\
  \bibinfo {author} {\bibfnamefont {H.}~\bibnamefont {Toda}},\ }\href@noop {}
  {\bibfield  {journal} {\bibinfo  {journal} {Int. J. Fatigue}\ }\textbf
  {\bibinfo {volume} {29}},\ \bibinfo {pages} {1039} (\bibinfo {year}
  {2007})}\BibitemShut {NoStop}%
\bibitem [{\citenamefont {Carter}\ \emph {et~al.}(1976)\citenamefont {Carter},
  \citenamefont {Hayes},\ and\ \citenamefont {Schurman}}]{carter}%
  \BibitemOpen
  \bibfield  {author} {\bibinfo {author} {\bibfnamefont {D.}~\bibnamefont
  {Carter}}, \bibinfo {author} {\bibfnamefont {W.~C.}\ \bibnamefont {Hayes}}, \
  and\ \bibinfo {author} {\bibfnamefont {D.~J.}\ \bibnamefont {Schurman}},\
  }\href@noop {} {\bibfield  {journal} {\bibinfo  {journal} {J. Biomech.}\
  }\textbf {\bibinfo {volume} {9}},\ \bibinfo {pages} {211} (\bibinfo {year}
  {1976})}\BibitemShut {NoStop}%
\bibitem [{\citenamefont {Li}\ \emph {et~al.}(2013)\citenamefont {Li},
  \citenamefont {Demirci},\ and\ \citenamefont {Silberschmidt}}]{li2013}%
  \BibitemOpen
  \bibfield  {author} {\bibinfo {author} {\bibfnamefont {S.}~\bibnamefont
  {Li}}, \bibinfo {author} {\bibfnamefont {E.}~\bibnamefont {Demirci}}, \ and\
  \bibinfo {author} {\bibfnamefont {V.~V.}\ \bibnamefont {Silberschmidt}},\
  }\href@noop {} {\bibfield  {journal} {\bibinfo  {journal} {J. Mech. Behav.
  Biomed. Mater.}\ }\textbf {\bibinfo {volume} {21}},\ \bibinfo {pages} {109}
  (\bibinfo {year} {2013})}\BibitemShut {NoStop}%
\bibitem [{\citenamefont {Mayya}\ \emph {et~al.}(2013)\citenamefont {Mayya},
  \citenamefont {Banerjee},\ and\ \citenamefont {Rajesh}}]{scirep}%
  \BibitemOpen
  \bibfield  {author} {\bibinfo {author} {\bibfnamefont {A.}~\bibnamefont
  {Mayya}}, \bibinfo {author} {\bibfnamefont {A.}~\bibnamefont {Banerjee}}, \
  and\ \bibinfo {author} {\bibfnamefont {R.}~\bibnamefont {Rajesh}},\
  }\href@noop {} {\bibfield  {journal} {\bibinfo  {journal} {Sci. Rep.}\
  }\textbf {\bibinfo {volume} {3}} (\bibinfo {year} {2013})}\BibitemShut
  {NoStop}%
\bibitem [{\citenamefont {Mayya}\ \emph
  {et~al.}(2016{\natexlab{a}})\citenamefont {Mayya}, \citenamefont {Banerjee},\
  and\ \citenamefont {Rajesh}}]{msec}%
  \BibitemOpen
  \bibfield  {author} {\bibinfo {author} {\bibfnamefont {A.}~\bibnamefont
  {Mayya}}, \bibinfo {author} {\bibfnamefont {A.}~\bibnamefont {Banerjee}}, \
  and\ \bibinfo {author} {\bibfnamefont {R.}~\bibnamefont {Rajesh}},\
  }\href@noop {} {\bibfield  {journal} {\bibinfo  {journal} {Mater. Sci. Eng.,
  C}\ }\textbf {\bibinfo {volume} {59}},\ \bibinfo {pages} {454} (\bibinfo
  {year} {2016}{\natexlab{a}})}\BibitemShut {NoStop}%
\bibitem [{\citenamefont {Ba{\v{z}}ant}(2004)}]{bazant2004}%
  \BibitemOpen
  \bibfield  {author} {\bibinfo {author} {\bibfnamefont {Z.~P.}\ \bibnamefont
  {Ba{\v{z}}ant}},\ }\href@noop {} {\bibfield  {journal} {\bibinfo  {journal}
  {Proc. Natl. Acad. Sci. U.S.A.}\ }\textbf {\bibinfo {volume} {101}},\
  \bibinfo {pages} {13400} (\bibinfo {year} {2004})}\BibitemShut {NoStop}%
\bibitem [{\citenamefont {Budyn}\ and\ \citenamefont {Hoc}(2010)}]{hoc}%
  \BibitemOpen
  \bibfield  {author} {\bibinfo {author} {\bibfnamefont {E.}~\bibnamefont
  {Budyn}}\ and\ \bibinfo {author} {\bibfnamefont {T.}~\bibnamefont {Hoc}},\
  }\href@noop {} {\bibfield  {journal} {\bibinfo  {journal} {Int. J. Numer.
  Meth. Eng.}\ }\textbf {\bibinfo {volume} {82}},\ \bibinfo {pages} {940}
  (\bibinfo {year} {2010})}\BibitemShut {NoStop}%
\bibitem [{\citenamefont {Abdel-Wahab}\ \emph {et~al.}(2012)\citenamefont
  {Abdel-Wahab}, \citenamefont {Maligno},\ and\ \citenamefont
  {Silberschmidt}}]{wahab}%
  \BibitemOpen
  \bibfield  {author} {\bibinfo {author} {\bibfnamefont {A.~A.}\ \bibnamefont
  {Abdel-Wahab}}, \bibinfo {author} {\bibfnamefont {A.~R.}\ \bibnamefont
  {Maligno}}, \ and\ \bibinfo {author} {\bibfnamefont {V.~V.}\ \bibnamefont
  {Silberschmidt}},\ }\href@noop {} {\bibfield  {journal} {\bibinfo  {journal}
  {Comput. Mater. Sci.}\ }\textbf {\bibinfo {volume} {52}},\ \bibinfo {pages}
  {128} (\bibinfo {year} {2012})}\BibitemShut {NoStop}%
\bibitem [{\citenamefont {Brown}\ \emph {et~al.}(2014)\citenamefont {Brown},
  \citenamefont {Tarsuslugil}, \citenamefont {Wijayathunga},\ and\
  \citenamefont {Wilcox}}]{brown2014}%
  \BibitemOpen
  \bibfield  {author} {\bibinfo {author} {\bibfnamefont {K.~R.}\ \bibnamefont
  {Brown}}, \bibinfo {author} {\bibfnamefont {S.}~\bibnamefont {Tarsuslugil}},
  \bibinfo {author} {\bibfnamefont {V.}~\bibnamefont {Wijayathunga}}, \ and\
  \bibinfo {author} {\bibfnamefont {R.}~\bibnamefont {Wilcox}},\ }\href@noop {}
  {\bibfield  {journal} {\bibinfo  {journal} {J. R. Soc. Interface}\ }\textbf
  {\bibinfo {volume} {11}},\ \bibinfo {pages} {20140186} (\bibinfo {year}
  {2014})}\BibitemShut {NoStop}%
\bibitem [{\citenamefont {Alava}\ \emph {et~al.}(2006)\citenamefont {Alava},
  \citenamefont {Nukala},\ and\ \citenamefont {Zapperi}}]{alava}%
  \BibitemOpen
  \bibfield  {author} {\bibinfo {author} {\bibfnamefont {M.~J.}\ \bibnamefont
  {Alava}}, \bibinfo {author} {\bibfnamefont {P.~K.}\ \bibnamefont {Nukala}}, \
  and\ \bibinfo {author} {\bibfnamefont {S.}~\bibnamefont {Zapperi}},\
  }\href@noop {} {\bibfield  {journal} {\bibinfo  {journal} {Adv. Phys.}\
  }\textbf {\bibinfo {volume} {55}},\ \bibinfo {pages} {349} (\bibinfo {year}
  {2006})}\BibitemShut {NoStop}%
\bibitem [{\citenamefont {Curtin}\ and\ \citenamefont {Scher}(1990)}]{curtin}%
  \BibitemOpen
  \bibfield  {author} {\bibinfo {author} {\bibfnamefont {W.}~\bibnamefont
  {Curtin}}\ and\ \bibinfo {author} {\bibfnamefont {H.}~\bibnamefont {Scher}},\
  }\href@noop {} {\bibfield  {journal} {\bibinfo  {journal} {J. Mater. Res.}\
  }\textbf {\bibinfo {volume} {5}},\ \bibinfo {pages} {535} (\bibinfo {year}
  {1990})}\BibitemShut {NoStop}%
\bibitem [{\citenamefont {Ray}(2006)}]{ray}%
  \BibitemOpen
  \bibfield  {author} {\bibinfo {author} {\bibfnamefont {P.}~\bibnamefont
  {Ray}},\ }\href@noop {} {\bibfield  {journal} {\bibinfo  {journal} {Comput.
  Mater. Sci.}\ }\textbf {\bibinfo {volume} {37}},\ \bibinfo {pages} {141}
  (\bibinfo {year} {2006})}\BibitemShut {NoStop}%
\bibitem [{\citenamefont {Moukarzel}\ and\ \citenamefont
  {Duxbury}(1994)}]{moukarzel}%
  \BibitemOpen
  \bibfield  {author} {\bibinfo {author} {\bibfnamefont {C.}~\bibnamefont
  {Moukarzel}}\ and\ \bibinfo {author} {\bibfnamefont {P.}~\bibnamefont
  {Duxbury}},\ }\href@noop {} {\bibfield  {journal} {\bibinfo  {journal} {J.
  Appl. Phys}\ }\textbf {\bibinfo {volume} {76}},\ \bibinfo {pages} {4086}
  (\bibinfo {year} {1994})}\BibitemShut {NoStop}%
\bibitem [{\citenamefont {Urabe}\ and\ \citenamefont {Takesue}(2010)}]{urabe}%
  \BibitemOpen
  \bibfield  {author} {\bibinfo {author} {\bibfnamefont {C.}~\bibnamefont
  {Urabe}}\ and\ \bibinfo {author} {\bibfnamefont {S.}~\bibnamefont
  {Takesue}},\ }\href@noop {} {\bibfield  {journal} {\bibinfo  {journal} {Phys.
  Rev. E}\ }\textbf {\bibinfo {volume} {82}},\ \bibinfo {pages} {016106}
  (\bibinfo {year} {2010})}\BibitemShut {NoStop}%
\bibitem [{\citenamefont {Boyina}\ \emph {et~al.}(2015)\citenamefont {Boyina},
  \citenamefont {Kirubakaran}, \citenamefont {Banerjee},\ and\ \citenamefont
  {Velmurugan}}]{dhatreyi}%
  \BibitemOpen
  \bibfield  {author} {\bibinfo {author} {\bibfnamefont {D.}~\bibnamefont
  {Boyina}}, \bibinfo {author} {\bibfnamefont {T.}~\bibnamefont {Kirubakaran}},
  \bibinfo {author} {\bibfnamefont {A.}~\bibnamefont {Banerjee}}, \ and\
  \bibinfo {author} {\bibfnamefont {R.}~\bibnamefont {Velmurugan}},\
  }\href@noop {} {\bibfield  {journal} {\bibinfo  {journal} {Mech. Mater.}\
  }\textbf {\bibinfo {volume} {91}},\ \bibinfo {pages} {64} (\bibinfo {year}
  {2015})}\BibitemShut {NoStop}%
\bibitem [{\citenamefont {Krajcinovic}\ \emph {et~al.}(1987)\citenamefont
  {Krajcinovic}, \citenamefont {Trafimow},\ and\ \citenamefont
  {Sumarac}}]{knovic}%
  \BibitemOpen
  \bibfield  {author} {\bibinfo {author} {\bibfnamefont {D.}~\bibnamefont
  {Krajcinovic}}, \bibinfo {author} {\bibfnamefont {J.}~\bibnamefont
  {Trafimow}}, \ and\ \bibinfo {author} {\bibfnamefont {D.}~\bibnamefont
  {Sumarac}},\ }\href@noop {} {\bibfield  {journal} {\bibinfo  {journal} {J.
  Biomech.}\ }\textbf {\bibinfo {volume} {20}},\ \bibinfo {pages} {779}
  (\bibinfo {year} {1987})}\BibitemShut {NoStop}%
\bibitem [{\citenamefont {Schwiedrzik}\ \emph {et~al.}(2014)\citenamefont
  {Schwiedrzik}, \citenamefont {Raghavan}, \citenamefont {B{\"u}rki},
  \citenamefont {LeNader}, \citenamefont {Wolfram}, \citenamefont {Michler},\
  and\ \citenamefont {Zysset}}]{schwiedrzik}%
  \BibitemOpen
  \bibfield  {author} {\bibinfo {author} {\bibfnamefont {J.}~\bibnamefont
  {Schwiedrzik}}, \bibinfo {author} {\bibfnamefont {R.}~\bibnamefont
  {Raghavan}}, \bibinfo {author} {\bibfnamefont {A.}~\bibnamefont {B{\"u}rki}},
  \bibinfo {author} {\bibfnamefont {V.}~\bibnamefont {LeNader}}, \bibinfo
  {author} {\bibfnamefont {U.}~\bibnamefont {Wolfram}}, \bibinfo {author}
  {\bibfnamefont {J.}~\bibnamefont {Michler}}, \ and\ \bibinfo {author}
  {\bibfnamefont {P.}~\bibnamefont {Zysset}},\ }\href@noop {} {\bibfield
  {journal} {\bibinfo  {journal} {Nat. Mater.}\ }\textbf {\bibinfo {volume}
  {13}},\ \bibinfo {pages} {740} (\bibinfo {year} {2014})}\BibitemShut
  {NoStop}%
\bibitem [{\citenamefont {Gunaratne}\ \emph {et~al.}(2002)\citenamefont
  {Gunaratne}, \citenamefont {Rajapaksa}, \citenamefont {Bassler},
  \citenamefont {Mohanty},\ and\ \citenamefont {Wimalawansa}}]{gunaratne}%
  \BibitemOpen
  \bibfield  {author} {\bibinfo {author} {\bibfnamefont {G.~H.}\ \bibnamefont
  {Gunaratne}}, \bibinfo {author} {\bibfnamefont {C.~S.}\ \bibnamefont
  {Rajapaksa}}, \bibinfo {author} {\bibfnamefont {K.~E.}\ \bibnamefont
  {Bassler}}, \bibinfo {author} {\bibfnamefont {K.~K.}\ \bibnamefont
  {Mohanty}}, \ and\ \bibinfo {author} {\bibfnamefont {S.~J.}\ \bibnamefont
  {Wimalawansa}},\ }\href@noop {} {\bibfield  {journal} {\bibinfo  {journal}
  {Phys. Rev. Lett.}\ }\textbf {\bibinfo {volume} {88}},\ \bibinfo {pages}
  {068101} (\bibinfo {year} {2002})}\BibitemShut {NoStop}%
\bibitem [{\citenamefont {Rajapakse}\ \emph {et~al.}(2004)\citenamefont
  {Rajapakse}, \citenamefont {Thomsen}, \citenamefont {Ortiz}, \citenamefont
  {Wimalawansa}, \citenamefont {Ebbesen}, \citenamefont {Mosekilde},\ and\
  \citenamefont {Gunaratne}}]{rajapakse}%
  \BibitemOpen
  \bibfield  {author} {\bibinfo {author} {\bibfnamefont {C.~S.}\ \bibnamefont
  {Rajapakse}}, \bibinfo {author} {\bibfnamefont {J.~S.}\ \bibnamefont
  {Thomsen}}, \bibinfo {author} {\bibfnamefont {J.~S.~E.}\ \bibnamefont
  {Ortiz}}, \bibinfo {author} {\bibfnamefont {S.~J.}\ \bibnamefont
  {Wimalawansa}}, \bibinfo {author} {\bibfnamefont {E.~N.}\ \bibnamefont
  {Ebbesen}}, \bibinfo {author} {\bibfnamefont {L.}~\bibnamefont {Mosekilde}},
  \ and\ \bibinfo {author} {\bibfnamefont {G.~H.}\ \bibnamefont {Gunaratne}},\
  }\href@noop {} {\bibfield  {journal} {\bibinfo  {journal} {J. Biomech.}\
  }\textbf {\bibinfo {volume} {37}},\ \bibinfo {pages} {1241} (\bibinfo {year}
  {2004})}\BibitemShut {NoStop}%
\bibitem [{\citenamefont {Tr{\c{e}}bacz}\ \emph {et~al.}(2013)\citenamefont
  {Tr{\c{e}}bacz}, \citenamefont {Zdunek}, \citenamefont {Cybulska},\ and\
  \citenamefont {Pieczywek}}]{trebacz}%
  \BibitemOpen
  \bibfield  {author} {\bibinfo {author} {\bibfnamefont {H.}~\bibnamefont
  {Tr{\c{e}}bacz}}, \bibinfo {author} {\bibfnamefont {A.}~\bibnamefont
  {Zdunek}}, \bibinfo {author} {\bibfnamefont {J.}~\bibnamefont {Cybulska}}, \
  and\ \bibinfo {author} {\bibfnamefont {P.}~\bibnamefont {Pieczywek}},\
  }\href@noop {} {\bibfield  {journal} {\bibinfo  {journal} {Australas. Phys.
  Eng. Sci. Med.}\ }\textbf {\bibinfo {volume} {36}},\ \bibinfo {pages} {43}
  (\bibinfo {year} {2013})}\BibitemShut {NoStop}%
\bibitem [{\citenamefont {Currey}(2002)}]{currey_book}%
  \BibitemOpen
  \bibfield  {author} {\bibinfo {author} {\bibfnamefont {J.~D.}\ \bibnamefont
  {Currey}},\ }\href@noop {} {\emph {\bibinfo {title} {Bones: structure and
  mechanics}}}\ (\bibinfo  {publisher} {Princeton university press},\ \bibinfo
  {year} {2002})\BibitemShut {NoStop}%
\bibitem [{\citenamefont {Mayya}\ \emph
  {et~al.}(2016{\natexlab{b}})\citenamefont {Mayya}, \citenamefont {Praveen},
  \citenamefont {Banerjee},\ and\ \citenamefont {Rajesh}}]{jrsoc}%
  \BibitemOpen
  \bibfield  {author} {\bibinfo {author} {\bibfnamefont {A.}~\bibnamefont
  {Mayya}}, \bibinfo {author} {\bibfnamefont {P.}~\bibnamefont {Praveen}},
  \bibinfo {author} {\bibfnamefont {A.}~\bibnamefont {Banerjee}}, \ and\
  \bibinfo {author} {\bibfnamefont {R.}~\bibnamefont {Rajesh}},\ }\href@noop {}
  {\bibfield  {journal} {\bibinfo  {journal} {J. Roy. Soc. Interface}\ }\textbf
  {\bibinfo {volume} {13}},\ \bibinfo {pages} {20160809} (\bibinfo {year}
  {2016}{\natexlab{b}})}\BibitemShut {NoStop}%
\bibitem [{\citenamefont {Landau}\ \emph {et~al.}(1986)\citenamefont {Landau},
  \citenamefont {Kosevich}, \citenamefont {Pitaevskii},\ and\ \citenamefont
  {Lifshitz}}]{landau1986}%
  \BibitemOpen
  \bibfield  {author} {\bibinfo {author} {\bibfnamefont {L.~D.}\ \bibnamefont
  {Landau}}, \bibinfo {author} {\bibfnamefont {A.}~\bibnamefont {Kosevich}},
  \bibinfo {author} {\bibfnamefont {L.~P.}\ \bibnamefont {Pitaevskii}}, \ and\
  \bibinfo {author} {\bibfnamefont {E.~M.}\ \bibnamefont {Lifshitz}},\
  }\href@noop {} {\emph {\bibinfo {title} {Theory of elasticity}}}\ (\bibinfo
  {publisher} {Butterworth},\ \bibinfo {year} {1986})\BibitemShut {NoStop}%
\bibitem [{\citenamefont {Monette}\ and\ \citenamefont
  {Anderson}(1994)}]{monette}%
  \BibitemOpen
  \bibfield  {author} {\bibinfo {author} {\bibfnamefont {L.}~\bibnamefont
  {Monette}}\ and\ \bibinfo {author} {\bibfnamefont {M.}~\bibnamefont
  {Anderson}},\ }\href@noop {} {\bibfield  {journal} {\bibinfo  {journal}
  {Modell. Simul. Mater. Sci. Eng.}\ }\textbf {\bibinfo {volume} {2}},\
  \bibinfo {pages} {53} (\bibinfo {year} {1994})}\BibitemShut {NoStop}%
\bibitem [{\citenamefont {Verlet}(1967)}]{verlet}%
  \BibitemOpen
  \bibfield  {author} {\bibinfo {author} {\bibfnamefont {L.}~\bibnamefont
  {Verlet}},\ }\href@noop {} {\bibfield  {journal} {\bibinfo  {journal} {Phys.
  Rev.}\ }\textbf {\bibinfo {volume} {159}},\ \bibinfo {pages} {98} (\bibinfo
  {year} {1967})}\BibitemShut {NoStop}%
\bibitem [{\citenamefont {Schaffler}\ and\ \citenamefont
  {Burr}(1988)}]{schaffler}%
  \BibitemOpen
  \bibfield  {author} {\bibinfo {author} {\bibfnamefont {M.~B.}\ \bibnamefont
  {Schaffler}}\ and\ \bibinfo {author} {\bibfnamefont {D.~B.}\ \bibnamefont
  {Burr}},\ }\href@noop {} {\bibfield  {journal} {\bibinfo  {journal} {J.
  Biomech.}\ }\textbf {\bibinfo {volume} {21}},\ \bibinfo {pages} {13}
  (\bibinfo {year} {1988})}\BibitemShut {NoStop}%
\bibitem [{\citenamefont {Currey}(1988)}]{currey_88}%
  \BibitemOpen
  \bibfield  {author} {\bibinfo {author} {\bibfnamefont {J.~D.}\ \bibnamefont
  {Currey}},\ }\href@noop {} {\bibfield  {journal} {\bibinfo  {journal} {J.
  Biomech.}\ }\textbf {\bibinfo {volume} {21}},\ \bibinfo {pages} {131}
  (\bibinfo {year} {1988})}\BibitemShut {NoStop}%
\bibitem [{\citenamefont {{\AA}kesson}\ \emph {et~al.}(1994)\citenamefont
  {{\AA}kesson}, \citenamefont {Grynpas}, \citenamefont {Hancock},
  \citenamefont {Odselius},\ and\ \citenamefont {Obrant}}]{aakesson}%
  \BibitemOpen
  \bibfield  {author} {\bibinfo {author} {\bibfnamefont {K.}~\bibnamefont
  {{\AA}kesson}}, \bibinfo {author} {\bibfnamefont {M.}~\bibnamefont
  {Grynpas}}, \bibinfo {author} {\bibfnamefont {R.}~\bibnamefont {Hancock}},
  \bibinfo {author} {\bibfnamefont {R.}~\bibnamefont {Odselius}}, \ and\
  \bibinfo {author} {\bibfnamefont {K.}~\bibnamefont {Obrant}},\ }\href@noop {}
  {\bibfield  {journal} {\bibinfo  {journal} {Calcif. Tissue Int.}\ }\textbf
  {\bibinfo {volume} {55}},\ \bibinfo {pages} {236} (\bibinfo {year}
  {1994})}\BibitemShut {NoStop}%
\bibitem [{\citenamefont {Bloebaum}\ \emph {et~al.}(1997)\citenamefont
  {Bloebaum}, \citenamefont {Skedros}, \citenamefont {Vajda}, \citenamefont
  {Bachus},\ and\ \citenamefont {Constantz}}]{bloebaum}%
  \BibitemOpen
  \bibfield  {author} {\bibinfo {author} {\bibfnamefont {R.~D.}\ \bibnamefont
  {Bloebaum}}, \bibinfo {author} {\bibfnamefont {J.}~\bibnamefont {Skedros}},
  \bibinfo {author} {\bibfnamefont {E.}~\bibnamefont {Vajda}}, \bibinfo
  {author} {\bibfnamefont {K.~N.}\ \bibnamefont {Bachus}}, \ and\ \bibinfo
  {author} {\bibfnamefont {B.}~\bibnamefont {Constantz}},\ }\href@noop {}
  {\bibfield  {journal} {\bibinfo  {journal} {Bone}\ }\textbf {\bibinfo
  {volume} {20}},\ \bibinfo {pages} {485} (\bibinfo {year} {1997})}\BibitemShut
  {NoStop}%
\bibitem [{\citenamefont {Petri}\ \emph {et~al.}(1994)\citenamefont {Petri},
  \citenamefont {Paparo}, \citenamefont {Vespignani}, \citenamefont {Alippi},\
  and\ \citenamefont {Costantini}}]{petri1994}%
  \BibitemOpen
  \bibfield  {author} {\bibinfo {author} {\bibfnamefont {A.}~\bibnamefont
  {Petri}}, \bibinfo {author} {\bibfnamefont {G.}~\bibnamefont {Paparo}},
  \bibinfo {author} {\bibfnamefont {A.}~\bibnamefont {Vespignani}}, \bibinfo
  {author} {\bibfnamefont {A.}~\bibnamefont {Alippi}}, \ and\ \bibinfo {author}
  {\bibfnamefont {M.}~\bibnamefont {Costantini}},\ }\href@noop {} {\bibfield
  {journal} {\bibinfo  {journal} {Phys. Rev. Lett.}\ }\textbf {\bibinfo
  {volume} {73}},\ \bibinfo {pages} {3423} (\bibinfo {year}
  {1994})}\BibitemShut {NoStop}%
\bibitem [{\citenamefont {Garcimartin}\ \emph {et~al.}(1997)\citenamefont
  {Garcimartin}, \citenamefont {Guarino}, \citenamefont {Bellon},\ and\
  \citenamefont {Ciliberto}}]{garcimartin1997}%
  \BibitemOpen
  \bibfield  {author} {\bibinfo {author} {\bibfnamefont {A.}~\bibnamefont
  {Garcimartin}}, \bibinfo {author} {\bibfnamefont {A.}~\bibnamefont
  {Guarino}}, \bibinfo {author} {\bibfnamefont {L.}~\bibnamefont {Bellon}}, \
  and\ \bibinfo {author} {\bibfnamefont {S.}~\bibnamefont {Ciliberto}},\
  }\href@noop {} {\bibfield  {journal} {\bibinfo  {journal} {Phys. Rev. Lett.}\
  }\textbf {\bibinfo {volume} {79}},\ \bibinfo {pages} {3202} (\bibinfo {year}
  {1997})}\BibitemShut {NoStop}%
\bibitem [{\citenamefont {Guarino}\ \emph {et~al.}(2002)\citenamefont
  {Guarino}, \citenamefont {Ciliberto}, \citenamefont {Garcimart{\i}n},
  \citenamefont {Zei},\ and\ \citenamefont {Scorretti}}]{guarino2002}%
  \BibitemOpen
  \bibfield  {author} {\bibinfo {author} {\bibfnamefont {A.}~\bibnamefont
  {Guarino}}, \bibinfo {author} {\bibfnamefont {S.}~\bibnamefont {Ciliberto}},
  \bibinfo {author} {\bibfnamefont {A.}~\bibnamefont {Garcimart{\i}n}},
  \bibinfo {author} {\bibfnamefont {M.}~\bibnamefont {Zei}}, \ and\ \bibinfo
  {author} {\bibfnamefont {R.}~\bibnamefont {Scorretti}},\ }\href@noop {}
  {\bibfield  {journal} {\bibinfo  {journal} {Eur. Phys. J. B}\ }\textbf
  {\bibinfo {volume} {26}},\ \bibinfo {pages} {141} (\bibinfo {year}
  {2002})}\BibitemShut {NoStop}%
\bibitem [{\citenamefont {Maes}\ \emph {et~al.}(1998)\citenamefont {Maes},
  \citenamefont {Van~Moffaert}, \citenamefont {Frederix},\ and\ \citenamefont
  {Strauven}}]{maes1998}%
  \BibitemOpen
  \bibfield  {author} {\bibinfo {author} {\bibfnamefont {C.}~\bibnamefont
  {Maes}}, \bibinfo {author} {\bibfnamefont {A.}~\bibnamefont {Van~Moffaert}},
  \bibinfo {author} {\bibfnamefont {H.}~\bibnamefont {Frederix}}, \ and\
  \bibinfo {author} {\bibfnamefont {H.}~\bibnamefont {Strauven}},\ }\href@noop
  {} {\bibfield  {journal} {\bibinfo  {journal} {Phys. Rev. B}\ }\textbf
  {\bibinfo {volume} {57}},\ \bibinfo {pages} {4987} (\bibinfo {year}
  {1998})}\BibitemShut {NoStop}%
\bibitem [{\citenamefont {Pradhan}\ \emph {et~al.}(2015)\citenamefont
  {Pradhan}, \citenamefont {Stroisz}, \citenamefont {Fj{\ae}r}, \citenamefont
  {Stenebr{\aa}ten}, \citenamefont {Lund},\ and\ \citenamefont
  {S{\o}nsteb{\o}}}]{pradhan2015}%
  \BibitemOpen
  \bibfield  {author} {\bibinfo {author} {\bibfnamefont {S.}~\bibnamefont
  {Pradhan}}, \bibinfo {author} {\bibfnamefont {A.~M.}\ \bibnamefont
  {Stroisz}}, \bibinfo {author} {\bibfnamefont {E.}~\bibnamefont {Fj{\ae}r}},
  \bibinfo {author} {\bibfnamefont {J.~F.}\ \bibnamefont {Stenebr{\aa}ten}},
  \bibinfo {author} {\bibfnamefont {H.~K.}\ \bibnamefont {Lund}}, \ and\
  \bibinfo {author} {\bibfnamefont {E.~F.}\ \bibnamefont {S{\o}nsteb{\o}}},\
  }\href@noop {} {\bibfield  {journal} {\bibinfo  {journal} {Rock Mech. Rock
  Eng.}\ }\textbf {\bibinfo {volume} {48}},\ \bibinfo {pages} {2529} (\bibinfo
  {year} {2015})}\BibitemShut {NoStop}%
\bibitem [{\citenamefont {Bar{\'o}}\ \emph {et~al.}(2016)\citenamefont
  {Bar{\'o}}, \citenamefont {Shyu}, \citenamefont {Pang}, \citenamefont
  {Jasiuk}, \citenamefont {Vives}, \citenamefont {Salje},\ and\ \citenamefont
  {Planes}}]{baro16pig}%
  \BibitemOpen
  \bibfield  {author} {\bibinfo {author} {\bibfnamefont {J.}~\bibnamefont
  {Bar{\'o}}}, \bibinfo {author} {\bibfnamefont {P.}~\bibnamefont {Shyu}},
  \bibinfo {author} {\bibfnamefont {S.}~\bibnamefont {Pang}}, \bibinfo {author}
  {\bibfnamefont {I.~M.}\ \bibnamefont {Jasiuk}}, \bibinfo {author}
  {\bibfnamefont {E.}~\bibnamefont {Vives}}, \bibinfo {author} {\bibfnamefont
  {E.~K.}\ \bibnamefont {Salje}}, \ and\ \bibinfo {author} {\bibfnamefont
  {A.}~\bibnamefont {Planes}},\ }\href@noop {} {\bibfield  {journal} {\bibinfo
  {journal} {Phys. Rev. E}\ }\textbf {\bibinfo {volume} {93}},\ \bibinfo
  {pages} {053001} (\bibinfo {year} {2016})}\BibitemShut {NoStop}%
\end{thebibliography}
%

\end{document}